\documentclass[a4paper,fleqn,usenatbib,useAMS]{mnras}

\usepackage{graphicx}	
\usepackage{amsmath}	
\usepackage{amssymb}	
\usepackage{multicol}        
\usepackage{bm}		
\usepackage{pdflscape}	

\usepackage[T1]{fontenc}
\usepackage{ae,aecompl}

\usepackage{newtxtext,newtxmath}
\usepackage{textcomp}

\title[AST3-2 data release]{Data Release of the AST3-2 Automatic Survey from Dome~A, Antarctica}

\author[Xu Yang. et al.]{Xu Yang$^{1,2}$, Yi Hu$^{1}$\thanks{Contact e-mail:
\href{mailto:huyi.naoc@gmail.com}{huyi.naoc@gmail.com}},Zhaohui Shang$^{1}$\thanks{Contact 
e-mail:\href{mailto:zshang@gmail.com}{zshang@gmail.com}},Bin Ma$^{3}$,  Michael C.B. 
Ashley$^{4}$,
\newauthor Xiangqun Cui$^{5,6}$, Fujia Du$^{5,6}$, Jianning Fu$^{7}$, Xuefei Gong$^{5,6}$, 
Bozhong Gu$^{5,6}$, 
\newauthor Peng Jiang$^{6,8}$, Xiaoyan Li$^{5,6}$, Zhengyang Li$^{5,6}$, Charling Tao$^{9,10}$,
Lifan Wang$^{6,11,12}$, 
\newauthor Lingzhe Xu$^{5}$, Shi-hai Yang$^{5,6}$, Ce Yu$^{13}$, Xiangyan Yuan$^{5,6}$, 
Ji-lin Zhou$^{14}$,
\newauthor Zhenxi Zhu$^{6,12}$
\\
\\$^{1}$National Astronomical Observatories, Chinese Academy of Sciences, Beijing 100101, China
\\$^{2}$University of Chinese Academy of Sciences, Beijing 100049, China
\\$^{3}$School of Physics and Astronomy, Sun Yat-Sen University, Zhuhai 519082, China
\\$^{4}$School of Physics, University of New South Wales, Sydney NSW 2052, Australia
\\$^{5}$Nanjing Institute of Astronomical Optics and Technology, Nanjing 210042, China
\\$^{6}$Chinese Center for Antarctic Astronomy, Nanjing 210008, China
\\$^{7}$Department of Astronomy, Beijing Normal University, Beijing, 100875, China
\\$^{8}$Key Laboratory for Polar Science, MNR, Polar Research Institute of China, Shanghai, 200136, China
\\$^{9}$Aix Marseille Univ., CNRS/IN2P3, CPPM, Marseille, France
\\$^{10}$Physics Department and Tsinghua Center for Astrophysics (THCA), Tsinghua University, 
Beijing, 100084, China
\\$^{11}$Department of Physics and Astronomy, Texas A\&M University, College Station, TX 77843, USA
\\$^{12}$Purple Mountain Observatory, Chinese Academy of Sciences, Nanjing 210008, China
\\$^{13}$College of Intelligence and Computing, Tianjin University, Tianjin 300350, China
\\$^{14}$School of Astronomy and Space Science, Nanjing University, Nanjing 210023, China}

\date{Accepted XXX. Received YYY; in original form ZZZ}

\pubyear{2023}

\begin{document}
\label{firstpage}
\pagerange{\pageref{firstpage}--\pageref{lastpage}}
\maketitle

\begin{abstract}
AST3-2 is the second of the three Antarctic Survey Telescopes, aimed at wide-field 
time-domain optical astronomy.  It is located at Dome A, Antarctica, which is by many measures 
the best optical astronomy site on the Earth's surface.  Here 
we present the data from the AST3-2 automatic survey in 2016 and the photometry 
results.  The median 5$\sigma$ limiting magnitude in $i$-band is 17.8 mag and the 
light curve precision is 4 mmag for bright stars.  The data release includes photometry for
over 7~million stars, from which over 3,500 variable stars were detected, 
with 70 of them newly discovered.  We classify these new variables into different 
types by combining their light curve features with stellar properties from surveys 
such as StarHorse.
\end{abstract}




\section{Introduction}
Time-domain astronomy has led to many astronomical discoveries through exploring the 
variability of astronomical objects over time.  Transient targets such as supernovae 
(SNe), gamma-ray bursts, and tidal disruption events (TDEs) give valuable insights in 
astronomy and fundamental physics.  Many survey projects have been undertaken to search 
for variable sources by repeatedly scanning selected sky areas.  Deep surveys over wide 
areas of sky require specialized telescopes such as the Large Binocular Telescope 
(LBT; \citealt{hill2000}) and the Large Synoptic Survey Telescope (LSST; \citealt{lsst2008}), 
and results from such surveys will doubtless make revolutionary discoveries in coming years.
High cadence is also important for time-domain surveys when searching for transients 
such as exoplanets, rapidly-changing objects, and short-term events.  The Wide Angle 
Search for Planets (WASP; \citealt{pollacco2006}) consortium has discovered numerous 
exoplanets with its high cadence.  The Zwicky Transient Facility (ZTF; \citealt{ztf2019}) 
has discovered over 3,000 supernovae from its first year of operations with a cadence as
rapid as 3 days.  

The Antarctic plateau is an ideal site for ground-based time-domain astronomy with its 
long clear polar nights that can provide long-term continuous observing time as well as 
other excellent observing conditions (\citealt{storey2005,storey2007,ashley2013}).  
The clean air can minimize the scattering of light, the cold air is good for 
infrared observations due to the low thermal background, and the stable atmosphere 
provides remarkably good seeing.

As the highest location on the Antarctic ice cap, Dome~A was first reached by the 21st 
CHInese National Antarctic Research Expedition (CHINARE) in 2005.  It is also the place 
where the Chinese Kunlun station was established.  
Many site testing studies have been conducted here during the past decade, and the results 
have confirmed that Dome~A is an excellent site for astronomical observations.  A complete 
summary of the astronomy-related work at Dome~A can be found in \citet{shang2020}.  We 
present some important results briefly below.

The Chinese Small Telescope ARray (CSTAR) showed that the median 
$i$-band sky background of moonless clear nights is 20.5 mag arcsec$^{-2}$ \citep{zou2010}.  
The KunLun Cloud and Aurora Monitor (KLCAM) showed that the nighttime clear sky rate is 83 per 
cent, which is better than most ground-based sites \citep{yang2021}.  Moreover, 
the Surface layer NOn-Doppler Acoustic Radar (SNODAR; \citealt{bonner2010}) showed a 
very shallow atmospheric turbulent boundary layer at Dome~A, with a median thickness of 
only 13.9\,m.  The multilayer Kunlun Automated Weather Station (KLAWS) showed that a 
temperature inversion often occurs near the ground, which leads to a stable atmosphere 
where cooler air is trapped under warmer air \citep{hu2014,hu2019}.  The results from SNODAR 
and KLAWS suggest that extremely good seeing is relatively easy to obtain at 
Dome~A since the telescope only has to be above the shallow turbulent boundary layer to 
achieve free-atmosphere conditions. This is impractical at traditional observatory sites 
where the boundary layer is typically many hundreds of metres above the ground.  
In 2019, the two KunLun Differential Image Motion Monitors (KL-DIMMs) directly confirmed 
these ideas by measuring the seeing at Dome~A from an 8\,m tall tower.  Superb night-time 
seeing as good as 0.13\arcsec was recorded.  The median free-atmosphere seeing was 
0.31\arcsec and the KL-DIMMs reached the free atmosphere from the 8m tower 31\% of the 
time \citep{ma2020a}.  In summary, the studies described above have demonstrated that by 
many measures Dome~A has the best optical observational conditions from the Earth's surface.

With such exceptional observing conditions, telescopes were planned and constructed 
to operate at Dome~A for time-domain astronomy.  The first-generation optical 
telescope, CSTAR, was installed in 2008 January \citep{yuan2008,zhou2010}.  
It observed a 20 deg$^{2}$ sky area centred at the South Celestial Pole with four co-aligned
14.5cm telescopes.  CSTAR obtained data for three years and has contributed to many
studies on stellar variability \citep{wang2011,yang2015,zong2015,liang2016,oelkers2016}. 
The three Antarctic Survey Telescopes (AST3; \citealt{cui2008}) were later planned 
as the second-generation optical telescopes at Dome~A, with larger apertures and the ability 
to point and track over the sky, as opposed to CSTAR's conservative engineering approach 
of having a fixed altitude.  

The first AST3 telescope (AST3-1) was installed at Dome~A in 2012 by the 28th CHINARE.  
AST3-1 surveyed a sky area of roughly 2000 deg$^{2}$ and the data have been released 
\citep{ma2018}.  AST3-1 also monitored some specific sky regions such as the Large and 
Small Magellanic Clouds.  These data were used for research on exoplanets and 
variable stars.  For example, AST3-1 detected about 500 variable stars around the 
Galactic disk centre, with 339 of them being newly discovered \citep{wang2017a}.  

The AST3 telescopes were originally conceived as multi-band survey telescopes operating 
together, but the goal has not been achieved due to various logistic difficulties, such 
as the required amount of electrical power.  
The second AST3 telescope (AST3-2) was installed in 2015 by the 31st CHINARE.  This 
work is based on the data from AST3-2.  The third AST3 (AST3-3) has been constructed 
and will be equipped with a K-dark infrared camera \citep{burton2016,li2016}.

Here we present the data and photometry from the AST3-2 sky survey as well as an analysis 
of the light curves.  We first present the basic design of AST3-2 in section \ref{sec:2} and 
go on to discuss the survey parameters and operational strategy
in section \ref{sec:3}.  In section \ref{sec:4} we discuss the
data reduction process and results.  In section \ref{sec:5} we present the light curves, 
the result of period searches, and the classification of objects.  The overall statistics 
of the catalogue and data access 
are discussed in section \ref{sec:6}.  Finally, we summarize the results in section \ref{sec:7}.  
 
\section{Instrument}
\label{sec:2}

The details of the AST3 system have been presented in previous works 
\citep{yuan2010,yuan2012,yuan2014}.  Here we briefly describe the basic features 
of the AST3-2, the second telescope of AST3.  

AST3-2 has the same modified Schmidt optical design as the \mbox{AST3-1}.  It has 
a 680mm primary mirror, an entrance pupil diameter of 500mm, a 3.73 f-ratio, and 
an SDSS $i$ filter.  The AST3 telescopes were designed specially to work in the harsh 
environment of Dome~A where the ambient temperature in the observation season ranges 
from $-80\degr$C to $-50\degr$C. The AST3 telescopes and the mounting system 
were built with low thermal expansion materials such as Invar to minimize the thermal 
effects.  This design enables the AST3-2 to work in extremely low temperatures, but we still had 
occasional problems with gears being stuck or jammed by ice.  To cope with optical element 
frosting problems that are common in Antarctica, a defrosting system was designed with an 
indium-tin-oxide (ITO) coating on the entrance aperture to the telescope and a warm blower 
inside the tube.  However, in the first year of operation, the frosting problem on the 
first surface was not completely solved.  The ITO coating was sometimes insufficient to 
defrost the ice and the blow heater had to work frequently, resulting 
in significant tube seeing and poor image quality.  To solve this problem, an external 
defrosting blower system was installed in front of the telescope in 2016. 

\begin{figure}
\includegraphics[width=\columnwidth]{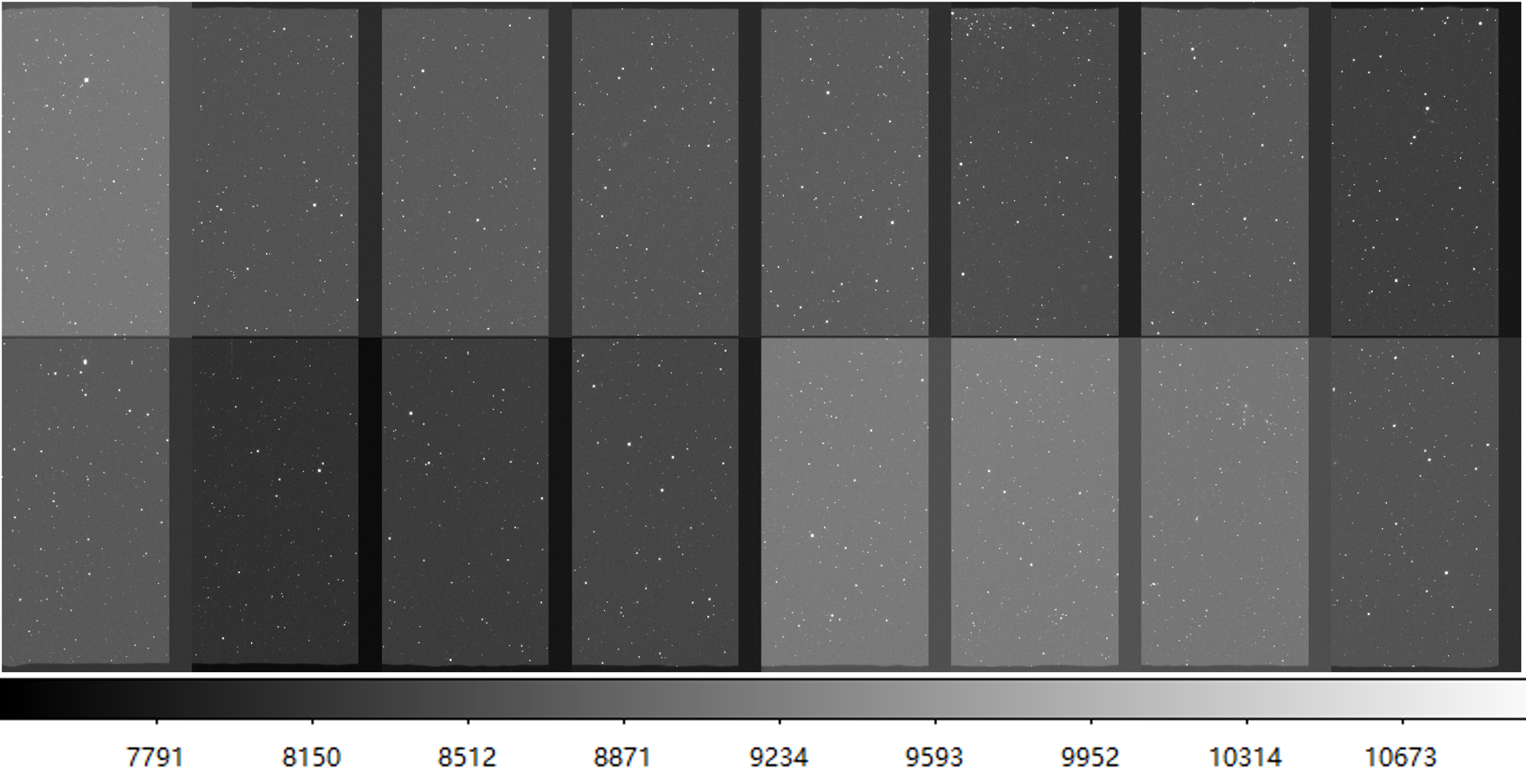}
\caption{
An example of a raw image that is taken from the survey fields by
AST3-2.  There are 16 readout channels with different bias levels and
overscan regions. The lower 8 channels are read out towards the bottom of the CCD, 
and the upper 8 channels are read out towards the top.  Each of the readouts has an 
area of 1500 pixels $\times$ 2660 pixels including overscan.  The overscan regions 
have 180 columns on the right of each readout and 20 horizontal rows in the
middle of the image.
}
\label{fig:raw-image}
\end{figure}

AST3-2 is equipped with a 10K $\times$ 10K STA1600FT CCD with a pixel 
size of 9\micron.  There are 16 read-out channels for the CCD to reduce the 
read-out time, which is 2.5s in fast read-out mode and 40s in slow read-out mode.  
To prevent shutter failure in cold weather, the camera works without a mechanical 
shutter, instead relying on frame-transfer 
mode and dedicating half of the CCD area to a buffer that is not exposed to light. 
The astronomically usable area of the CCD is therefore 10K~$\times$~5K pixels, with 
a scale of ~1\arcsec/pixel over a FOV of 2.93\degr $\times$ 1.47\degr.  
Since the CCD camera is installed inside the telescope tube, it also faced some 
heat dissipation problems, causing the CCD to often operate at temperatures as warm as 
-50$\degr$C to -40$\degr$C, leading to a significant dark current.  
Since we could not take dark frames on-site and the previously-taken laboratory dark 
images have different patterns, a new method was developed to derive a dark frame 
from the science images and will be discussed in section \ref{sec:darkcurrent}.  
There was also a problem with the AST3 CCD in that the photon transfer curve became 
non-linear at a level around 25000~ADU, leading to the brighter-fatter effect \citep{ma2014b}.  
Fig.~\ref{fig:raw-image} shows a raw image taken by AST3-2.  Detailed laboratory tests of the 
CCD performance can be found in \citet{ma2012} and \citet{shang2012}.

The AST3-2 is powered by the PLATeau Observatory for Dome~A (PLATO-A;
 \citealt{ashley2010}).  PLATO-A is a self-contained automated platform
providing an average power of 1kW for at least 1 year.  It also provides
Internet access through the Iridium satellite constellation.  The hardware and
software of the control, operation, and data system (CODS) of AST3-2
were designed to be responsible for the automated sky survey
\citep{shang2012,hu2016a,shang2016,ma2020b}.  The CODS consists of the
main control system, the data storage array, and the pipeline system.
To ensure the success of the sky survey, we developed the CODS to be stable
and reliable under the conditions of low power availability (1 kW), low data 
bandwidth (a maximum of about 2 GB over the course of the year), and the unattended 
situation in the harsh winter of Dome~A.  The supporting software provides a
fully automatic survey control and a real-time data processing pipeline on-site.

\section{Observations and Data}
\label{sec:3}
The observing season at Dome~A starts in mid-March when the Sun reaches 13 degrees 
below the horizon, i.e., at the end of twilight \citep{zou2010}.  The automated and 
unattended AST3 sky survey strategy was designed to optimize the available observing 
time and was realized with a survey scheduler in the CODS software \citep{liu2018}. The 
scheduler provides three different survey modes depending on the scientific 
requirements.  The SN survey mode mainly focuses on a survey for SNe and other 
transients, the exoplanet search mode aims at discovering and monitoring short-period 
exoplanets, and an additional special mode mainly targets the follow-up of transients.  

\begin{figure}
\includegraphics[width=\columnwidth]{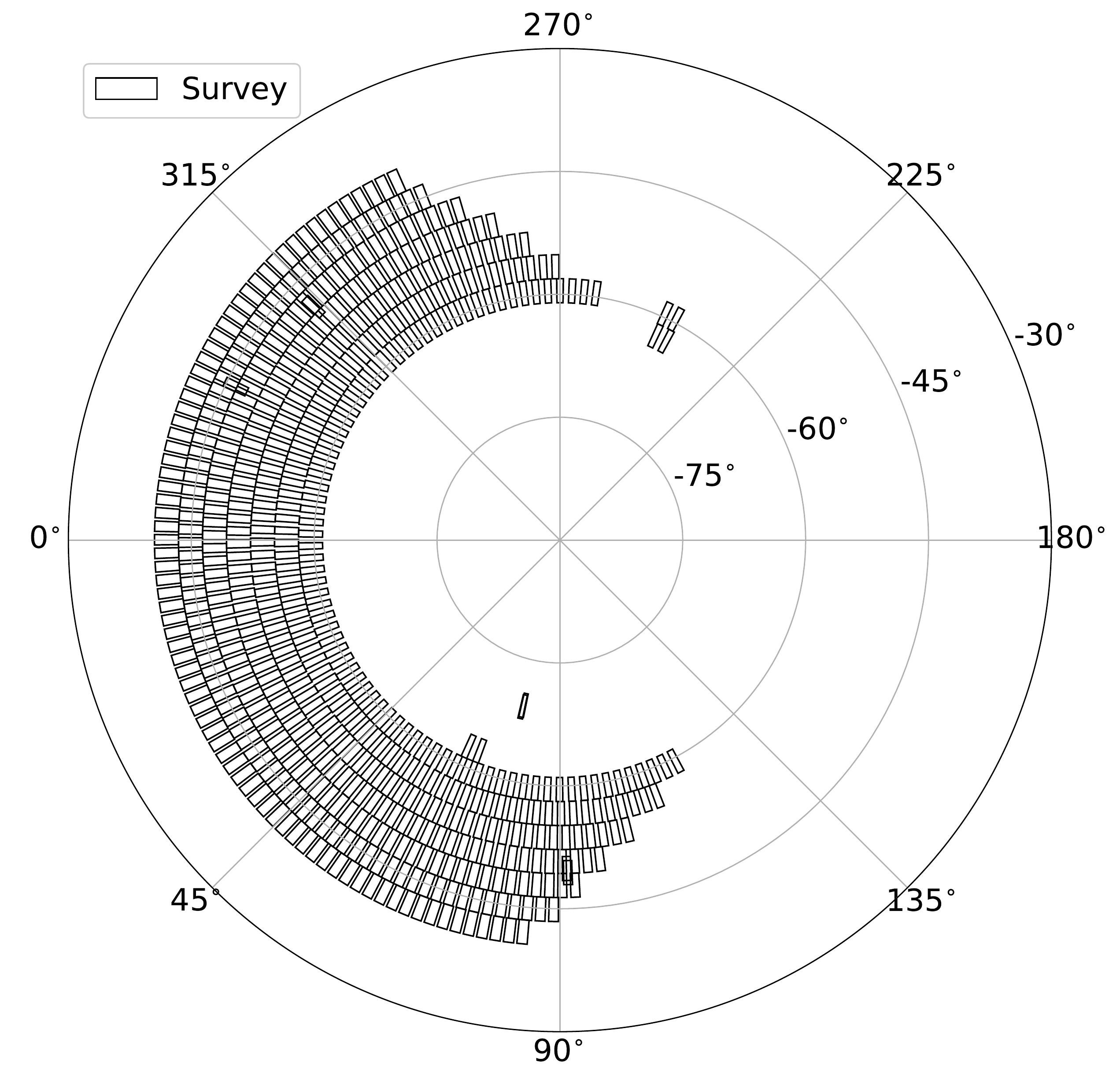}
\caption{The sky coverage of survey observations from AST3-2 in 2016. 
Each rectangle region 
is a target sky field based on the survey scheduler.
}
\label{fig:coverage}
\end{figure}

Following twilight, the AST3-2 was initially dedicated to the SN survey mode, lasting from 
2016 March~24 to May~16, at which point the long continuous polar night began and the 
survey switched to exoplanet mode.  The SN survey was designed for the early discovery 
of SNe as well as other transients, and for time-domain astronomy of variable 
stars.  It surveyed sky areas of 2200~deg$^{2}$, covering 565~fields with about 
30~visits each in a cadence of a half to a few days based on the fraction of dark 
time within a day.  Fig.~\ref{fig:coverage} shows the sky coverage of this survey.  
The real-time pipeline from CODS performed onsite data reduction and sent the SN or 
other transient candidates back to China for further confirmation and follow-up observations.  
For example, the real-time pipeline discovered the SN 2016ccp \citep{hu2016b} 
and the Type IIP SN 2017fbq \citep{wang2017b}.  During the test observations in Mohe, 
China, the AST3-2 recorded the SN 2014J in M82\citep{ma2014c} and discovered the 
type Ia SN 2014M \citep{ma2014d}.  The real-time pipeline is also 
capable of detecting other variables such as dwarf novae \citep{ma2016}, although 
most of the variables were not reported in the real-time pipeline.  So in this work, 
we mainly use the SN~survey mode data when the hard disks were physically returned from 
Dome~A to obtain the photometric catalogue and light curves of other variables.

The AST3-2 exoplanet project is named the CHinese Exoplanet Searching Program from 
Antarctica (CHESPA).  To search for short-period exoplanets rapidly and continuously, 
the exoplanet search mode started during the period of continuous dark polar nights: 
from May~16 to June~22.  The exoplanet search covered a smaller sky area than the SN 
survey, with 10 to 20~fields in each target region.  The target region during 2016 
contained 10~adjacent fields from the southern continuous viewing zone of TESS 
\citep{ricker2009}.  This part of the data has been analysed in previous works 
\citep{zhang2019a,zhang2019b,liang2020}.

Finally, a special mode was designed for the rapid follow-up of observations of 
interesting transients from the AST3-2 SN or exoplanet surveys, or from surveys 
by other telescopes.  This mode has the highest priority.  When an 
interesting target triggers the alert, it will pause other observations and resume 
them after the special observation is finished.  In 2017, AST3-2 successfully detected 
the first optical counterpart of the gravitational wave source GW170817 \citep{hu2017}.

\section{Data Reduction}
\label{sec:4}

The 2016 data of AST3-2 was retrieved by the 33rd CHINARE.  We focus on the 
SN survey data for this work.  First, we carried out the image corrections for 
CCD image pre-processing, cross-talk, image trimming, overscan, dark current, 
flat-field, and an unusual diagonal stripey noise described below.  Then we performed photometric 
and astrometric calibration to obtain the source catalogue.  Finally, we 
cross-matched the catalogues to obtain the light curves.  Details of the data 
reduction process are discussed in the subsections below.  

\subsection{Preprocessing}
\subsubsection{Image trimming and overscan subtraction}

The AST3 raw image has 12000~$\times$~5300 pixels including overscan regions 
and is divided into 16 channels with a size of 1500~$\times$~2650 each.  
As described in section \ref{sec:2}, the AST3 CCD works in frame-transfer 
mode, which means it does not have a shutter.  Since the zero-second exposure 
is not a true zero because the frame transfer period takes time, photons 
would be gathered in the 0s bias frame when there is no shutter.  This design 
makes it hard to take a bias frame on-site.  Instead, we used the overscan 
regions to remove the effect of the bias voltage.  As Fig.~\ref{fig:raw-image} 
shows, the overscan regions are the right 180 columns of each channel 
and 20 rows in the middle of the full raw image.  

Because the top and bottom rows of the CCD are insensitive to light,
we removed another 80~rows each from the top and bottom of the CCD
full images.  After overscan correction and image trimming, the final raw
images have a size of 10560~$\times$~5120 pixels.

\subsubsection{Dark current subtraction}
\label{sec:darkcurrent}

As described in section \ref{sec:2}, the CCD temperature was not very
stable and could be above $-50^{\circ}$C sometimes, making the dark
current non-negligible.  Moreover, the laboratory dark images had
different patterns and were not usable for dark correction 
in practice.  Additionally, we
could not take dark frames on site because the CCD does not have a
shutter and the AST3 was unattended for at least one year.
Therefore, a new
method was developed to derive the dark frame from scientific images
to solve this problem and it had been successfully applied to the AST3-1
images \citep{ma2014a,ma2018}.  Here we briefly describe this method
and how we utilized it in the AST3-2 preprocessing.

\begin{figure}
\includegraphics[width=\columnwidth]{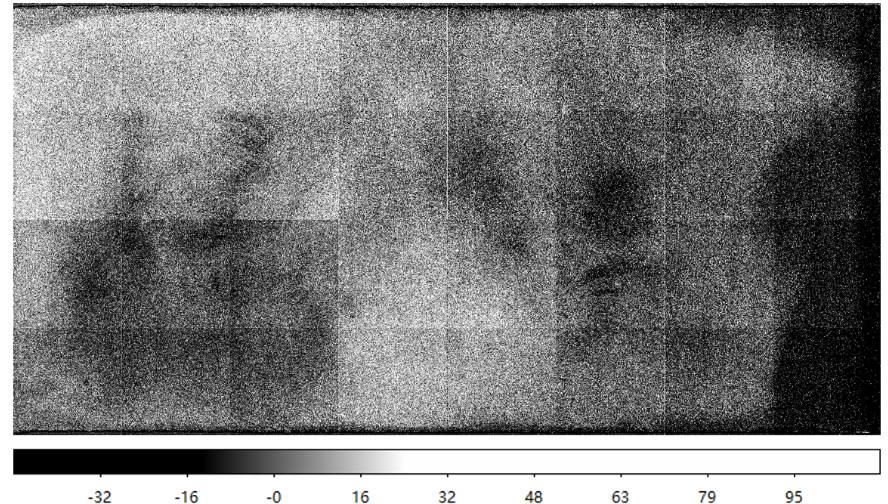}
\caption{The dark frame generated from observation. 
The difference between bright and dark 
regions is obvious.}
\label{fig:dark}
\end{figure}

The brightness $I$ of a pixel $(x, y)$ can be described as follow:
\begin{equation}
    I(x,y) = S_T + D(T) + \Delta d(T, x, y), \label{con:eq1}
\end{equation}
where $S_T$ is the sky background, $D(T)$ is the median dark current level at 
temperature $T$, and $\Delta d(T,x,y)$ is the deviation from the median dark 
current in pixel $(x, y)$ at temperature $T$.  The stars can be ignored by a 
median algorithm if we combine large numbers of images from different sky fields.  
For a single image, $D(T)$ can be considered constant.  Also, the sky brightness 
can be considered a constant because it is spatially flat enough after twilight 
\citep{yang2017}.  The first two terms on the right-hand side of the equation 
(\ref{con:eq1}) can be considered constant for a single image.  To derive the 
distribution of the deviation from the median dark current level $\Delta d(T,x,y)$, 
we need to take two scientific images that were taken at the same temperature but 
with different sky brightnesses.  By scaling the two images to an equivalent median 
level and subtracting one from another, we can derive a $\Delta d(T,x,y)$ image at a 
specific temperature $T$.  We repeated this process for different pairs of images 
at the same $T$ and combined the dark images to construct a master dark image for 
the specific temperature $T$.  Fig.~ \ref{fig:dark} shows the master dark image 
derived from the 2016 observations.

For different temperatures, the dark current level doubles as the temperature 
increases every 7.3$^{\circ}$ for the AST3 CCD between $-80^{\circ}$ and $-40^{\circ}$ 
\citep{ma2012}.  We used this relation to scale the master dark image to different 
temperatures and correct the dark current for all images.  

\subsubsection{Flat field correction}
During the beginning of the observing season, we took numerous twilight sky images 
and produced a master flat-field image.  The large FOV of AST3 led to a non-uniform 
large-scale gradient of the twilight images, which varied with the Sun elevation and 
angular distance from the field.  The method of brightness gradient correction was 
studied in \citet{wei2014}.  Two-dimensional fitting was applied to each flat 
image to correct the brightness gradient.  Finally, we median combined the corrected 
flat images to construct a master flat-field image.  After the correction, the mean 
root-mean-square (RMS) of the master flat was far below 1 per cent.

\subsubsection{Cross-talk and stripey noise corrections}
\label{sec:stripey}

\begin{figure}
\includegraphics[width=\columnwidth]{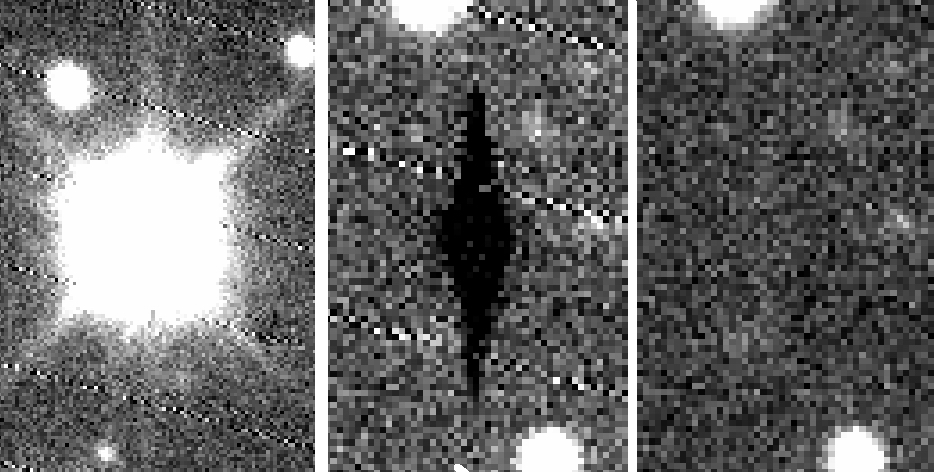}
\caption{An example of the cross-talk effect in the AST3 raw image.  The left 
panel shows a saturated star in one channel that would cause the cross-talk 
effect in other channels.  The middle panel shows the mirror pixels where the 
saturated pixels are in another CCD channel.  The stripey noise can also be 
seen in this area.  The right panel shows the same image as in the middle panel 
but after the cross-talk effect and the stripey noise are corrected.
}
\label{fig:crosstalk}
\end{figure}

Due to the simultaneous CCD readouts, when one amplifier reads a saturated pixel, 
other amplifiers will be affected.  There are significant CCD cross-talk effects 
in the raw images.  As Fig.~\ref{fig:crosstalk} shows, when one saturated pixel 
is read in one readout channel, the other 15 channels will have a negative ghost 
image at the exact position of the saturated pixel presenting as a dark spot.  
To remove the effect, we initially planned to locate all the saturated pixels, find 
the position of the related ghost pixels, and add the appropriate negative values back.  
However, the unsaturated pixels around the saturated ones also have cross-talk 
effects, making the ghost images hard to locate.  So we developed a method to 
correct the cross-talk effect during the correction for the stripey noise, described below.  

\begin{figure}
\includegraphics[width=\columnwidth]{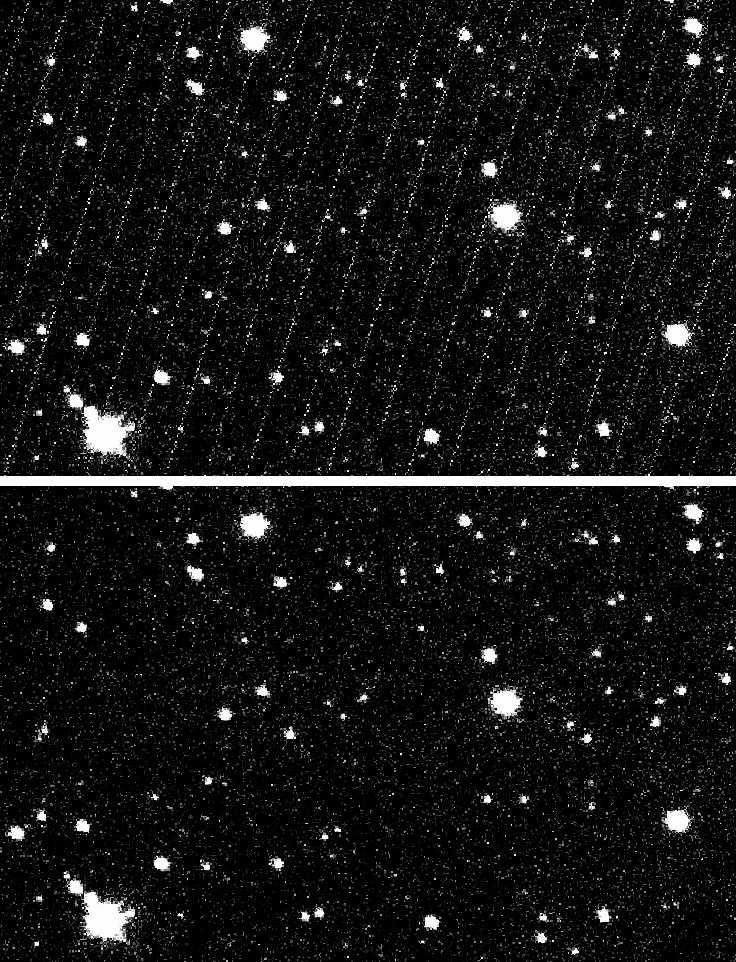}
\caption{Upper: An example of the stripey noise in the AST3 raw image due to a break 
in a cable shield.  Lower: The same image area but with the noise removed through the 
method discussed in section \ref{sec:stripey}.  
}
\label{fig:strip-noise}
\end{figure}

As Fig.~\ref{fig:strip-noise} shows, the raw images of AST3 in 2016 
have shown an unusual kind of stripey noise.  After careful investigation,
we found that the diagonal stripes were due to electromagnetic
interference at 16 kHz caused by a broken ground shield in the cables for the 
telescope's DC motor drives.  Because
this noise lies in exactly the same positions in each of the 16 CCD channels, 
and is extremely reproducible, for
each channel we constructed a filtered image from the other 15
channels by median combining the star-removed images of single
channels.  By subtracting from each channel the filtered image, we can
remove the stripey noise to the point where it is not detectable, as
Fig.~\ref{fig:strip-noise} shows.  
The pattern of the noise is similar to the cross-talk effect, which
also lies at the same position of different readout channels.  So, the above
method also helped to correct the cross-talk problem.  

\subsection{Photometry and astrometry}
\subsubsection{Photometry}

\begin{figure}
\includegraphics[width=\columnwidth]{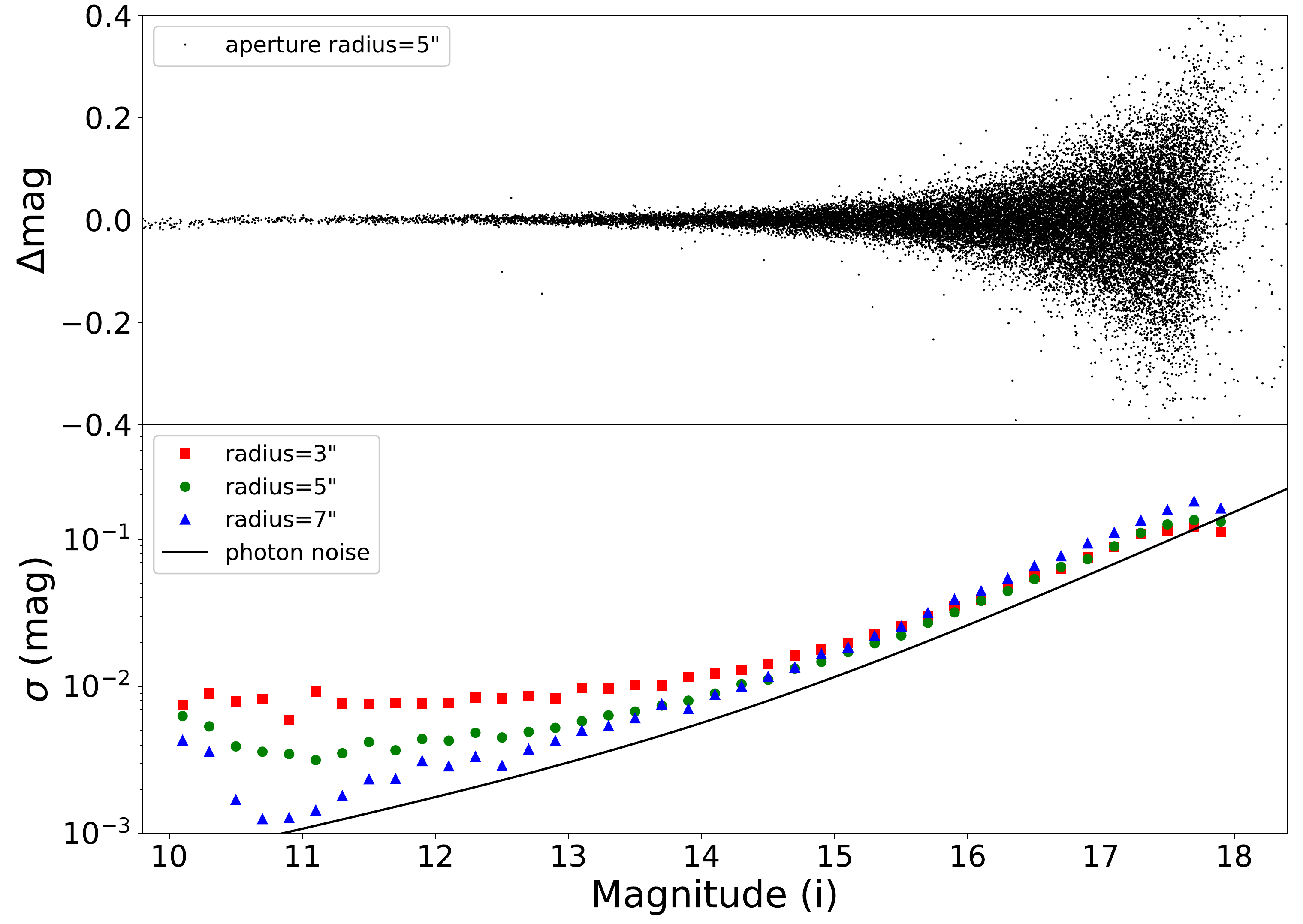}
\caption{
The upper panel presents the magnitude differences between two consecutive exposures 
b160505.000122 and b160505.000123 as a function of $i$-band magnitude measured with a 
circular aperture of 3\arcsec in radius. The lower panel presents the magnitude RMS 
calculated in 0.2 magnitude bins.  The different colours represent different circular 
apertures.  The solid lines are the expected error from photon noise of 5\arcsec radius.  
The trend of magnitude $\sigma$ changing with aperture goes contrary 
at different magnitude ends.  At the brighter end, the $\sigma$ is lower when the 
aperture radius is higher. It means for brighter stars larger apertures are appropriate 
when the photon noise of the star itself is dominant.  The stars brighter than 11 
magnitudes are saturated and have higher $\sigma$.  At the fainter end, the $\sigma$ 
is lower when the aperture radius is lower.  This is because the sky background dominated 
the noise for fainter stars and a smaller aperture is more suitable.  The number of 
stars is insufficient at the very faint end, thus the $\sigma$ is not real and seems to 
be smaller than the ideal photon noise.   
}
\label{fig:2image}
\end{figure}

We performed aperture photometry using the Source Extractor
\citep{bertin1996}.  Considering the changing full width at half
maximum (FWHM) of our images, we used multiple apertures to adapt to the varying
image quality.  The aperture radii were set to 3, 5, and 7 pixels.  Because 
the median FWHM of our data is
5 pixels, we set the default aperture radius as 5 pixels, or 5\arcsec
at our pixel scale of 1\arcsec/pixel.  An additional Kron-like
elliptical aperture magnitude MAG\_AUTO was adopted for galaxies.
Fig.~\ref{fig:2image} shows the photometric accuracy of two consecutive
images by comparing the magnitude differences between them.  

\subsubsection{Astrometry}

For astrometry, we used SCAMP to solve the World Coordinate System
\citep{bertin2006}.  We adopted the Position and Proper Motions
eXtended catalogue (PPMX) as a reference, which contains 440 sources
per deg$^{2}$ with the one-dimensional precision of 40 mas
\citep{ppmx2008}.  As a result, the external precision of our
astrometric calibration is 0.1\arcsec and the internal precision is
0.06\arcsec, in both Right Ascension (RA) and Declination (Dec.).

\subsubsection{Flux calibration}
We adopted the SkyMapper catalogue as the $i$-band magnitude reference for the flux 
calibration \citep{skymapper2018}.  The SkyMapper Southern Survey is a southern 
hemispheric survey carried out with the SkyMapper Telescope at Siding Spring Observatory 
in Australia. It covers an area of 17,200 deg$^{2}$ and the limiting magnitude reaches a 
depth of roughly 18 magnitudes in $uvgriz$ pass band.  

We first chose our best frame of each survey field for absolute 
calibration.  The ``best'' refers to the images that have the best image 
quality in one field considering the number of detected sources, background 
brightness, FWHM, and elongation.  Then we calculate the zero point in 
$i$-band magnitude for calibration.  We only chose the stars that are 
between 11 and 14 magnitudes in the $i$-band for calibration to 
balance high accuracy with a sufficient number of stars.  

After the absolute calibration, we used these calibrated images as 
references to relatively calibrate the other images of each survey 
field.  However, we found that the zero point changes with position 
and the cause still requires further investigation.  
To avoid large field non-uniformity of the zero point, we decide to do the 
flux calibration in each readout channel.  

\begin{figure*}
\includegraphics[width=\textwidth]{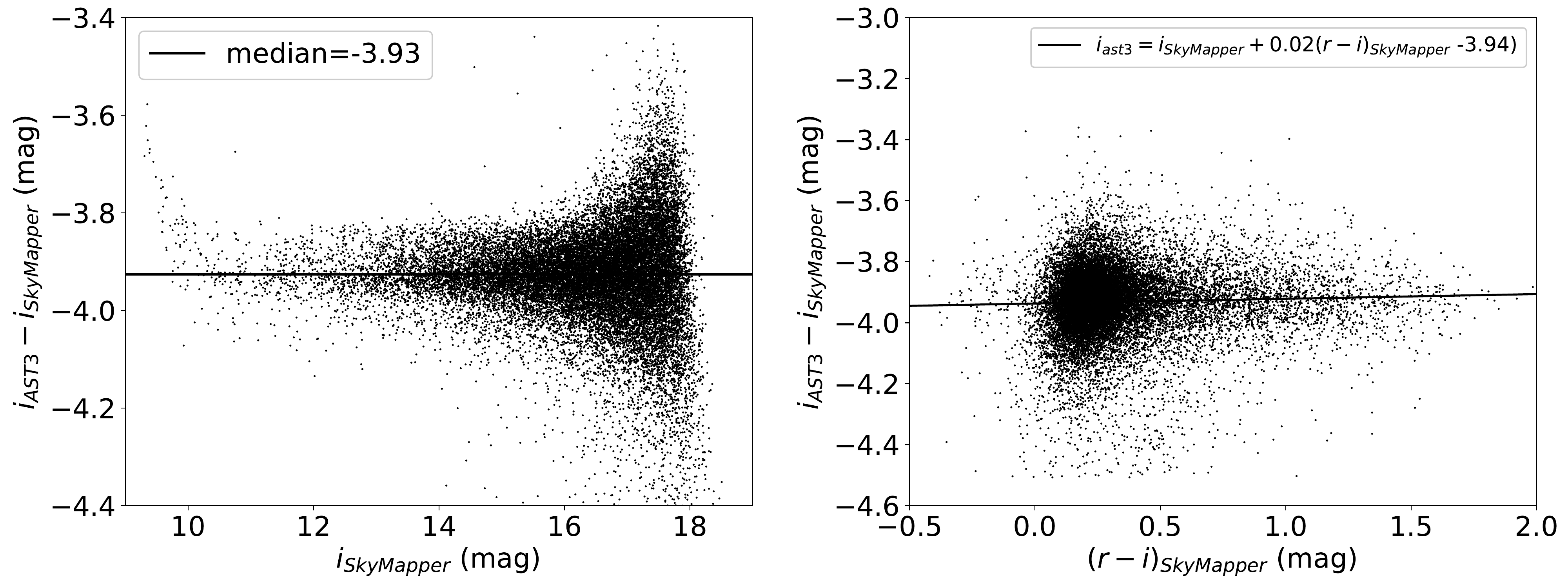}
\caption{Left panel: The magnitude difference between AST3-2 and SkyMapper in the AST3-2 
image b160505.000122. Right panel: The difference between the $i$-band catalogue of SkyMapper 
and AST3-2 versus the SkyMapper $r-i$ magnitude.
}
\label{fig:colour}
\end{figure*}

For the AST3-2 survey, we only have $i$-band data.  To investigate the
colour term, we compared the AST3-2 $i$-band data with SkyMapper $i$-
and $g$-band data as Fig.~\ref{fig:colour} shows.  The colour
coefficient is 0.02, much smaller than that for AST3-1 reported by
\citet{ma2018}.  However, we used a different reference catalogue from
AST3-1, which adopted the AAVSO Photometric All-Sky Survey catalogue
(APASS; \citet{henden2016}).  Our $i$-band magnitude matches
relatively well with the SkyMapper catalogue, but to compare with
other catalogues observed in the same band we need to be cautious.  

\subsection{Data quality}
\label{sec:dataquality}
\begin{figure*}
\includegraphics[width=\textwidth]{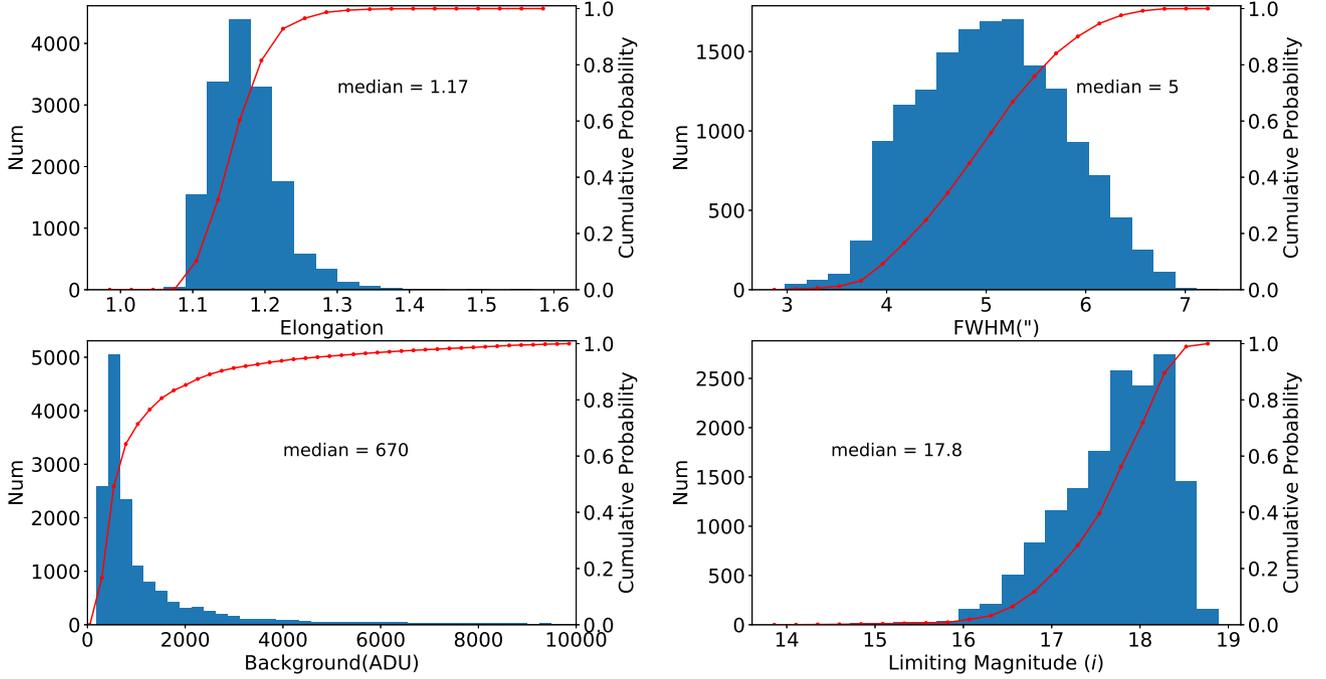}
\caption{The statistics of the star elongation, FWHM, sky background, and limiting 
magnitude of the data, which had median values of
1.17, 5\arcsec, 670~ADU, and 17.8~mag, respectively.  
}
\label{fig:quality}
\end{figure*}

Fig.~\ref{fig:quality} displays the distributions of data quality, 
showing the median value of elongation of $\sim$1.17, 
FWHM of 5\arcsec, background of 670~ADU, and 
limiting magnitude of 17.8~mag.  Some issues with tracking stability
AST3-2 led to the elongated star profiles.  We also see this 
problem from the range of FWHM, which varies from 3 to 7 arcseconds. 
Another cause of the wide FWHM distribution was the changing 
tube seeing.  In the extremely cold and high relative humidity conditions at Dome~A, 
there can be frost on the first surface of the optical system that reduces the 
transmission and changes the point-spread function through scattering.  
As described in section \ref{sec:2}, a heater and a blower were used to prevent 
the frosting problem, and the tube seeing would be unstable when they 
were working.  As a result, the limiting 
magnitude is not as good as that of the first AST3 telescope AST3-1 \citep{ma2018}.

\section{Stellar Variability and Statistics}
\label{sec:5}
\subsection{Time series}
\label{sec:timeseries}
\begin{figure}
\includegraphics[width=\columnwidth]{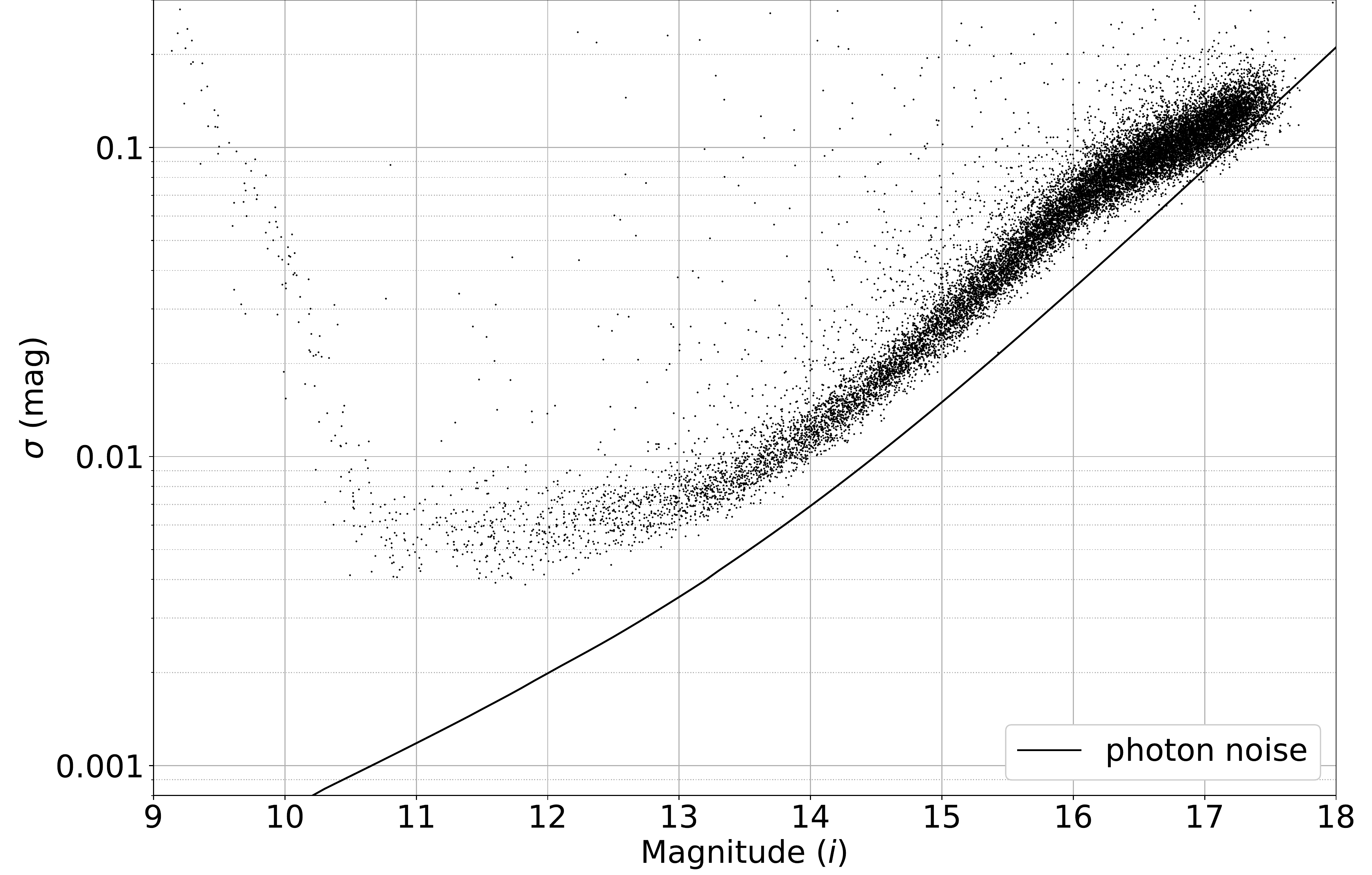}
\caption{The light curve rms as a function of magnitude.  Each data point 
represents a light curve from the region b160505.000122.
}
\label{fig:sigma}
\end{figure}

Images with poor quality were first excluded to ensure the quality of the light 
curves.  Such images could be due to heavy frost, or doubling of stars images from 
tracking problems.  We also excluded images with a background 
brightness larger than 10,000 ADU, median FWHM larger than 8\arcsec, 
fewer than 2000 stars, and median elongations larger 
than 2.  In this way, we excluded about 30 per cent of the images.  
We then cross-matched the targets in each field and obtained light 
curves.  Finally, an additional outlier elimination was performed to 
remove the false targets with obvious anomalous magnitudes and FWHMs.  
Fig.~\ref{fig:sigma} shows a typical light curve dispersion with an 
aperture radius of 5\arcsec. 

\subsection{Period search}
\label{sec:5.2}
On average, we observed each survey field 30 times during the year. 
Some of the targets might not be detected in some images 
due to poor image quality etc.  Thus, together with the image selections 
in section \ref{sec:timeseries},  for each target, the total number of 
epochs could be less than 30.  To analyse the stellar variability with 
enough detection and better image quality, we restricted ourselves to 
sky fields with more than 30 observations.  There are about one-third of the 
observations were exposed 3 times continuously, originally for image combination.  
Due to the tracking problem discussed in section \ref{sec:dataquality} and 
section \ref{sec:timeseries}, some of the multi-images would be excluded in the image 
selections and we tend not to combine them.  For the remaining multi-images, we did not 
count them as individual observations, but we used them as independent data points in 
light curve analysis.  Then, we rejected the targets that were only detected in less than 
50 per cent of the images.  Finally, in the period analysis, we chose the 
light curves with a significant variability of more than 2.5$\sigma$.  

\begin{figure}
\includegraphics[width=\columnwidth]{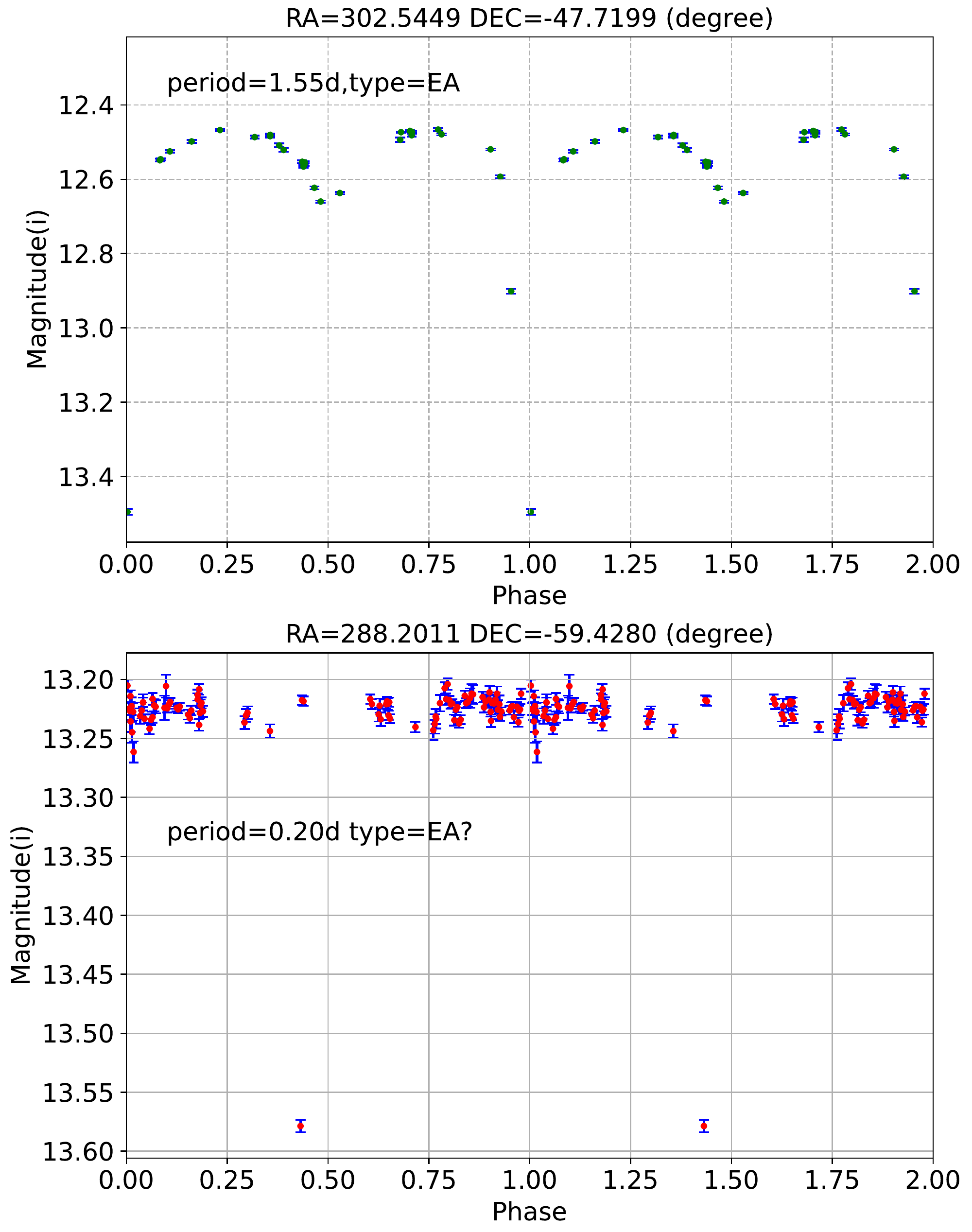}
\caption{Upper: An example light curve of an EA-type variable star folded in 2 phases.  
Lower: An example false signal showing an EA-type variable pattern.
}
\label{fig:lc-example}
\end{figure}

For our survey data, the sampling in the light curves with time is not uniform and 
thus we used the Lomb-Scargle (LS) method for period search \citep{lomb1976,scargle1982}. 
Light curves with a signal-to-noise ratio (SNR) larger than 5 are considered eligible 
candidates.  Then we cross-matched the candidate light curves with the International 
Variable Star Index (VSX; \citealt{watson2006}) and found 3,551 known variables.  For 
candidates that were not in the VSX catalogue, we visually inspected whether their 
periodicities were significant or not.  For candidates that were significantly variable 
and periodic, we then checked whether it was a false signal.  For example, 
Fig.\ref{fig:lc-example} shows a comparison of the true and false EA-type variable 
candidates.  The former is an EA-type variable candidate included in the VSX catalogue.  
The latter shows a similar light curve pattern but turned out to be a false signal 
affected by an outlier.  We manually excluded these kinds of false signals and we 
take the true ones as variable candidates.  In total, we found 70 new variables.  

\subsection{New variables}
For the newly discovered variable candidates, we tried to visually
classify them into different classes by their periods, amplitudes, and
light curve patterns.  We also obtained their effective temperature,
surface gravity, and metallicity from StarHorse \citep{starhorse2019} 
to help the classification.  Moreover, we obtained their B-V from the UCAC4 
\citep{ucac42013}, APASS9 \citep{henden2016}, NOMAD \citep{nomad2004}, 
and SPM4.0 \citep{spm2011} catalogues.  
However, due to insufficient observations, it was still hard to
classify them such as the example in section \ref{sec:5.2}.  
The insufficient observations at the minimum luminosity make it hard to classify.  
Many under-sampled candidates were excluded from the candidate list if they were 
not a known variable.

\begin{figure*}
\includegraphics[width=\textwidth]{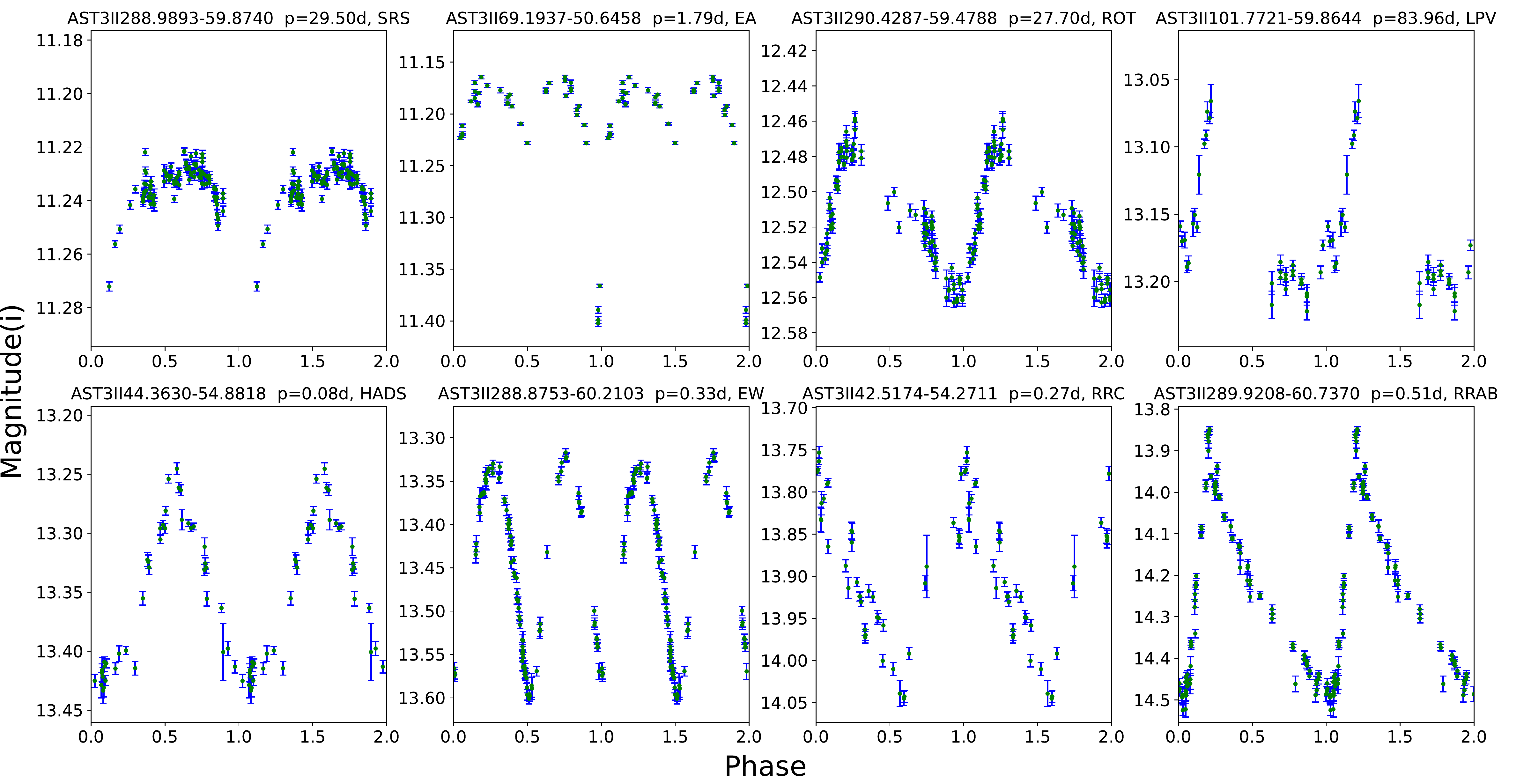}
\caption{Example light curves of the variables in different classes folded in 2 phases.  
Periods and types are marked on the light curves.
}
\label{fig:lc-class}
\end{figure*}

For this reason, we were only able to classify the candidates into 5
different classes.  There are 17 candidates classified as long-period
variables (LPV) either because they were observed in less than one
period or because we could hardly distinguish one periodic signal from
its light curve.  We found 5 candidates with Cepheid-like signals and we
classified them as pulsating stars (PUL).  We found 4 candidates as
eclipsing binary (EC) candidates by their periods and patterns.  
Of the remaining candidates, 24 of them have small amplitudes ($<0.1$) and long
periods (a few days to a dozen days), and they are likely to be
rotational variables (ROT).  The final 20 candidates have periods shorter
than 2 days and some of their periods are even shorter than 0.2 days.  Most of
these candidates have strange phase diagram patterns and we are not
sure whether they are real or a result of a lack of data points.  Under this
circumstance, we classified them as possible rotational variables (pROT).
Fig. \ref{fig:lc-class} shows the typical phased or time-series light curves of
each class.  
As mentioned in section \ref{sec:5.2}, when we try to classify the light curves, we only
consider the ones with 30 epochs or more to ensure there are enough
observations for a reliable period.  We can confirm some of
them that have obvious and distinctive light curve patterns.  But for many
variables that met the 30 epoch threshold, the absence of critical data points
in the light curves might lead to a false period, and an incorrect pattern in their phase
diagrams.  In such cases we erred on the side of not claiming them as newly discovered
variables.  

We cross-matched our variable candidates with the VSX catalogue version 2022-10-31.  
Interestingly, we initially used an earlier version of the VSX catalogue and our count of
new candidates was 126; 56 of these were listed in the latest version, which gave us 
the opportunity of comparing our classifications with VSX. The classifications agreed well, 
with disagreements mainly with LPVs and ECs.  Some stars were identified as rotational 
variables in the VSX that we classified as LPVs because we have relatively time coverage 
and we considered all the light curves with less than one period as LPVs.  As for the ECs 
in VSX, we classified some of them as ROTs or pROTs since we did not have enough critical 
data points to confirm them.  

\section{Data Availability}
\label{sec:6}
The AST3-2 data is available through the Chinese Astronomical Data Center 
(CADC)\footnote{\url{https://cstr.cn/11379.11.100669}}$^,$
\footnote{\url{https://doi.org/10.12149/100669}}.  
The data contains an $i$-band catalogue, a light curve catalogue, and preprocessed images.   

The $i$-band catalogue contains over 7~million sources with a median limiting 
magnitude of 17.8 mag.  For objects with multiple observations, we adopted their median 
positions and median magnitudes.  Table \ref{tab:surveyinfo} shows the Database Schema of 
the catalogue.  

Table \ref{tab:lcinfo} details the information in the light curve catalogue.  
The light curves are presented as time series and the catalogue 
contains information from every observation after quality filtering.  The periodic variables 
discussed in this work are also presented and listed in Appendix \ref{sec:appendix}.  

There are also 22576 images in the format of Flexible Image Transport System (FITS) 
presented in the data set.  These are the preprocessed FITS images discussed in 
section \ref{sec:4} with observing information such as date, exposure time, 
and WCS coordinates.  

\begin{table}
    \caption{AST3-2 survey catalogue Table Schema}
    \label{tab:surveyinfo}
    \centering
    \begin{tabular}{l|l}
    \hline
         Column Name  &  Description \\
    \hline
         ID & Source index \\
         RA & Right Ascension in J2000 (deg) \\
         Dec. & Declination in J2000 (deg) \\
         MAG & Median aperture magnitudes (mag) \\
         MAGERR & Standard deviation of magnitudes (mag) \\
         COUNT & Number of observations \\
    \hline
    \end{tabular}
\end{table}

\begin{table}
    \caption{AST3-2 light curve catalogue Table Schema}
    \label{tab:lcinfo}
    \centering
    \begin{tabular}{l|l}
    \hline
         Column Name  &  Description \\
    \hline
         DATE & The beginning time of observation in \\
              & ISO time \\
         MJD & The beginning time of observation in \\
             & Modified Julian date \\
         X & Windowed X position in CCD (pixel) \\
         Y & Windowed Y position in CCD (pixel) \\
         RA & Right Ascension in J2000 (deg) \\
         DEC & Declination in J2000 (deg) \\
         MAG & Aperture magnitudes in 5\arcsec radius (mag) \\
         MAGERR & Aperture magnitude errors in 5\arcsec radius \\
                & (mag) \\
         FLUX & Flux (ADU) \\
         FLUXERR & Flux error (ADU) \\
         MAG$\_$AUTO & Magnitude in Kron aperture (mag) \\
         MAGERR$\_$AUTO & Magnitude error in Kron aperture (mag) \\
         BACKGROUND & Background brightness (ADU) \\
         FWHM & Full width at half-maximum in Gaussian \\
              & profile (pixel) \\
         ELONGATION & Ratio of semi-major to semi-minor axis \\
         A & Semimajor axis length (pixel) \\
         B & Semiminor axis length (pixel) \\
         THETA & Position angle of semimajor axis (degrees \\
               & east from north) \\
         MAG$\_$3 & Aperture magnitudes in 3\arcsec radius (mag) \\
         MAGERR$\_$3 & Aperture magnitude errors in 3\arcsec radius \\
                & (mag) \\
         MAG$\_$7 & Aperture magnitudes in 7\arcsec radius (mag) \\
         MAGERR$\_$7 & Aperture magnitude errors in 7\arcsec radius \\
                 & (mag) \\
         
    \hline
    \end{tabular}
\end{table}

\section{Summary}
\label{sec:7}
The second AST3 telescope AST3-2 was deployed at Dome~A, Antarctica in
2015.  In 2016, it worked fully automatically on a sky survey for SNae and
semi-automatically on an exo-planet search.  In this work, we report on
the 2016~SN survey data observed between Mar.~23 and May~16.  We
surveyed 2200 deg$^2$ fields with about 30 visits each in a cadence of
a half to a few days.  After the raw data was retrieved, we
preprocessed the data, performed aperture photometry, calibrated the
magnitudes, obtained the light curves of the 565 sky fields, and
briefly studied the variability of the light curves.  In this
paper, we present the data release of the photometric data from the
AST3-2 SN survey in 2016.  It consists of 22000 scientific images,
7~million sources brighter than $i \sim $18 with photometry, 
astrometry, and light curves.

The 5$\sigma$ limiting magnitude of this dataset is 17.8 mag with 4 mmag 
precision in the light curves of bright stars.  The median FWHM, 
elongation, and background brightness are 5.0\arcsec, 1.17, and 
670~ADU, respectively.  We found 70 new 
variable candidates out of $\sim$ 3,500 variable stars.  We 
check the stellar properties from surveys such as StarHorse to 
help us classify these variables into 5 types.

\section*{Acknowledgements}
\addcontentsline{toc}{section}{Acknowledgements}
We thank the CHINARE for their great efforts in installing 
AST3-2, maintaining AST3-2 and PLATO-A, and retrieving data. 
This work has been supported by the National Natural 
Science Foundation of China under Grant Nos. 11873010, 11733007, 11673037, 
11403057, and 11403048, the Chinese Polar Environment Comprehensive 
Investigation and Assessment Programmes under grant No. 
CHINARE2016-02-03, and the National Basic Research Program of China 
(973 Program) under Grant No. 2013CB834900. PLATO-A is supported by the 
Australian Antarctic Division.  Data Publishing is supported by China National 
Astronomical Data Center (NADC), CAS Astronomical Data Center and Chinese Virtual 
Observatory (China-VO). 




\bibliographystyle{mnras}
\bibliography{ref} 




\appendix
\section{The Light curves of New Candidates}
\label{sec:appendix}
Fig.\ref{fig:lc-new} shows the phased new variable candidates, and Table \ref{tab:variablelist} 
shows the list of them.

\begin{figure*}
\includegraphics[width=\textwidth]{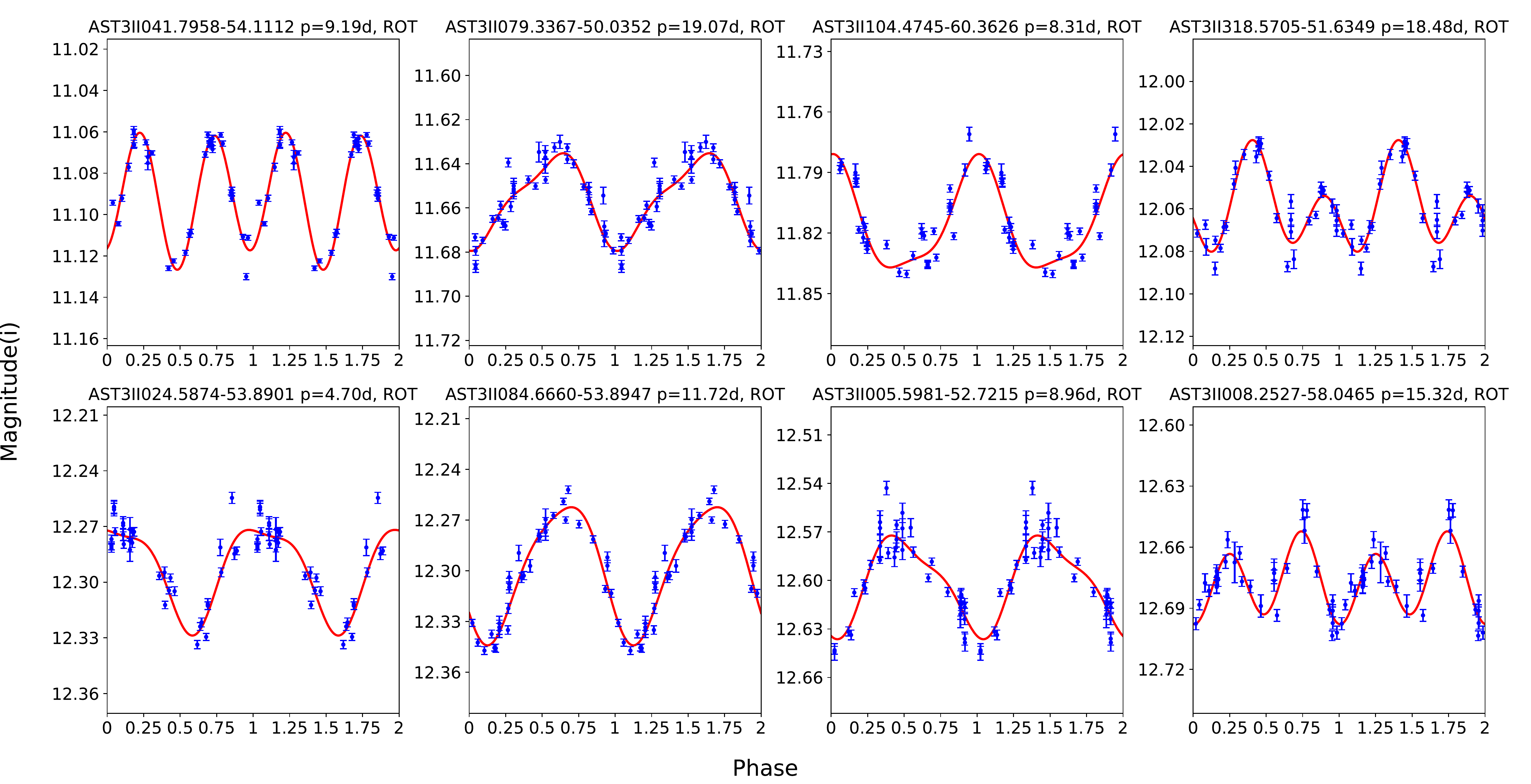}
\caption{Light curves of the new candidates in different classes folded in 2 phases 
except the LPVs show no apparent periods and are in the form of time series.  The 
dots with error bars are observed data and the red curve is the fitted line.  Periods 
and types are marked on the light curves.  The pulsating stars are marked as PUL and 
the possible rotational variables are marked as pROT.  
}
\label{fig:lc-new}
\end{figure*}

\begin{figure*}
\includegraphics[width=\textwidth]{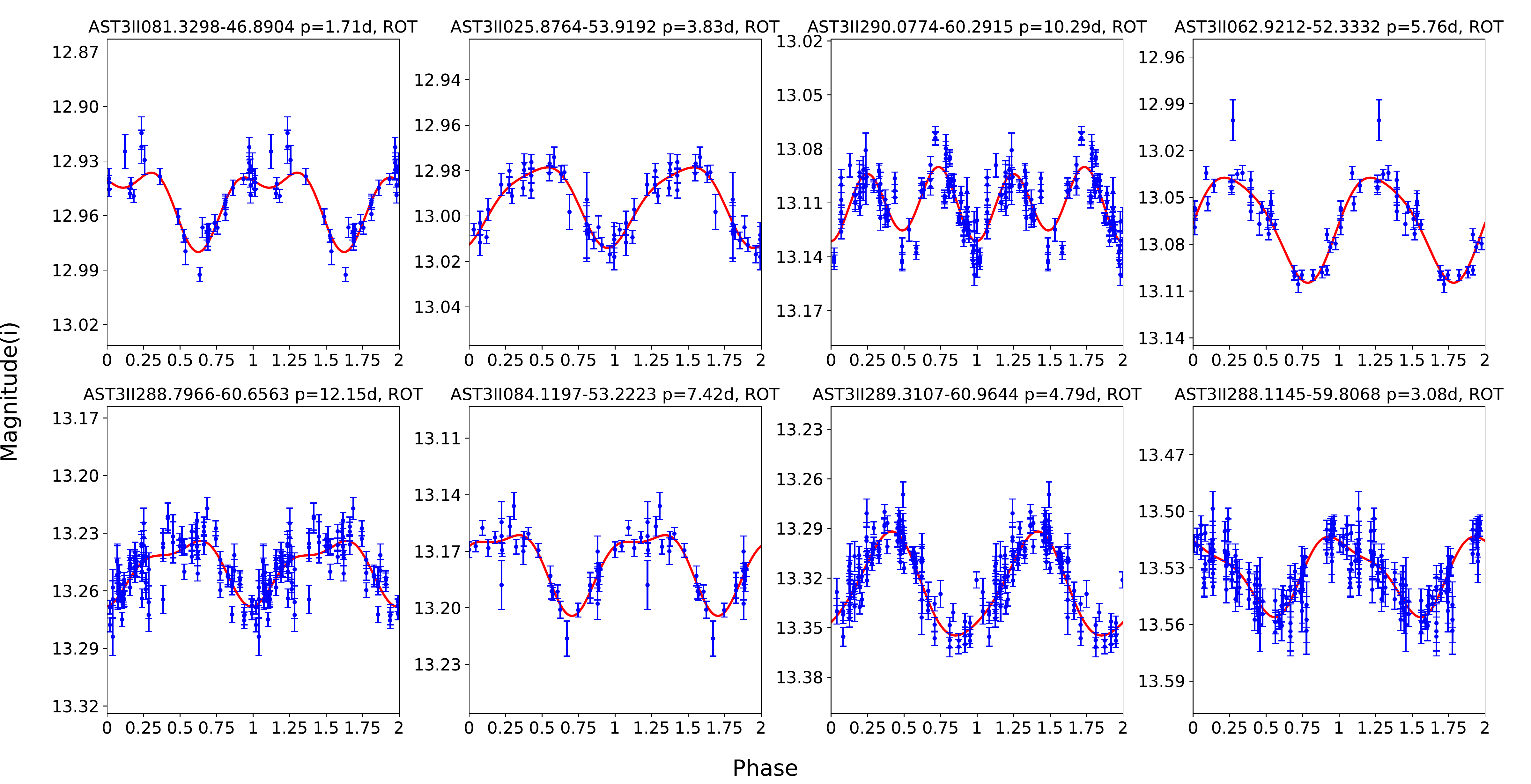}
\contcaption{Light curves of the new candidates in different classes folded in 2 phases 
except the LPVs show no apparent periods and are in the form of time series.  The 
dots with error bars are observed data and the red curve is the fitted line.  Periods 
and types are marked on the light curves.  The pulsating stars are marked as PUL and 
the possible rotational variables are marked as pROT.  
}
\label{fig:lc-new-2}
\end{figure*}

\begin{figure*}
\includegraphics[width=\textwidth]{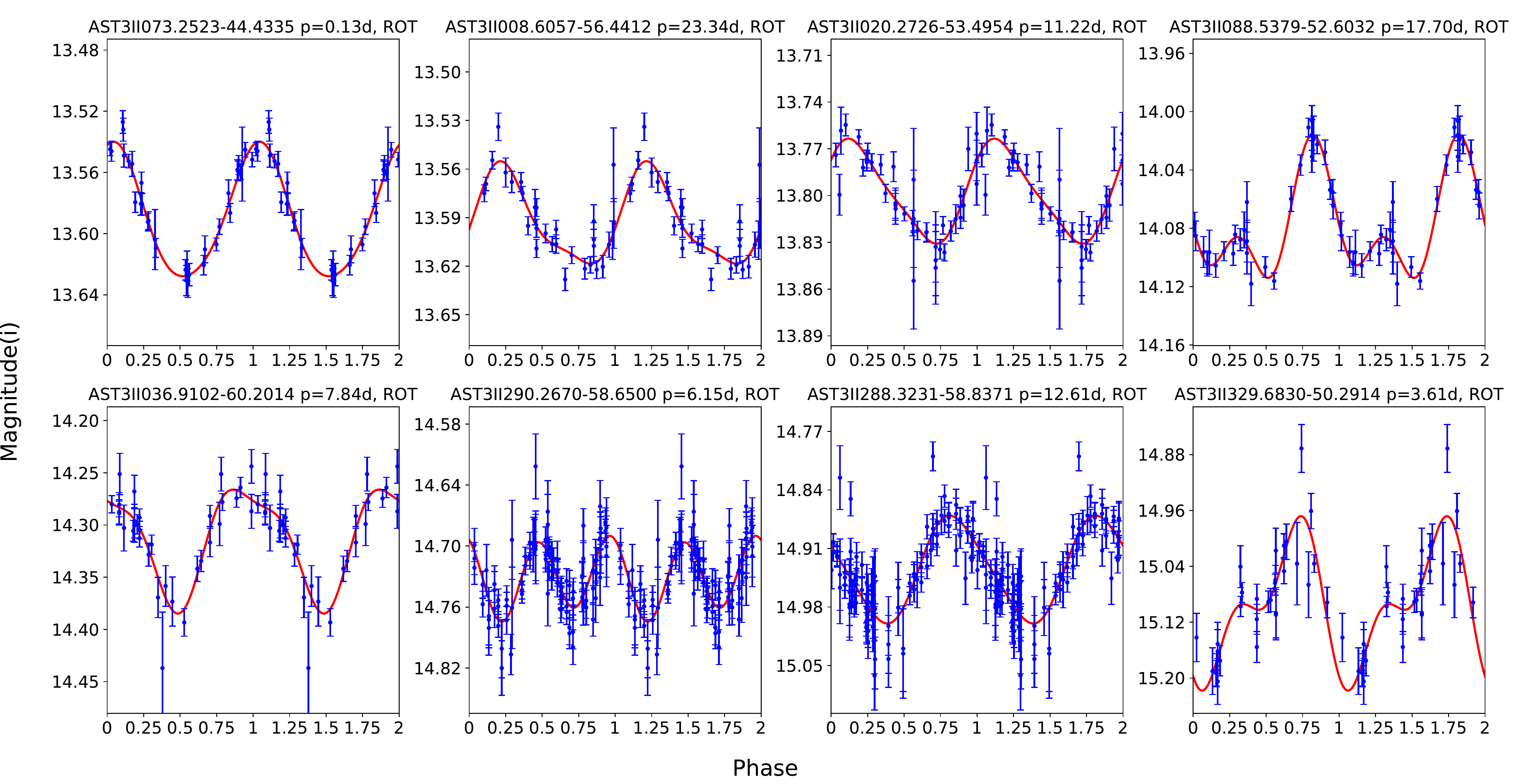}
\contcaption{Light curves of the new candidates in different classes folded in 2 phases 
except the LPVs show no apparent periods and are in the form of time series.  The 
dots with error bars are observed data and the red curve is the fitted line.  Periods 
and types are marked on the light curves.  The pulsating stars are marked as PUL and 
the possible rotational variables are marked as pROT.  
}
\label{fig:lc-new-3}
\end{figure*}

\begin{figure*}
\includegraphics[width=\textwidth]{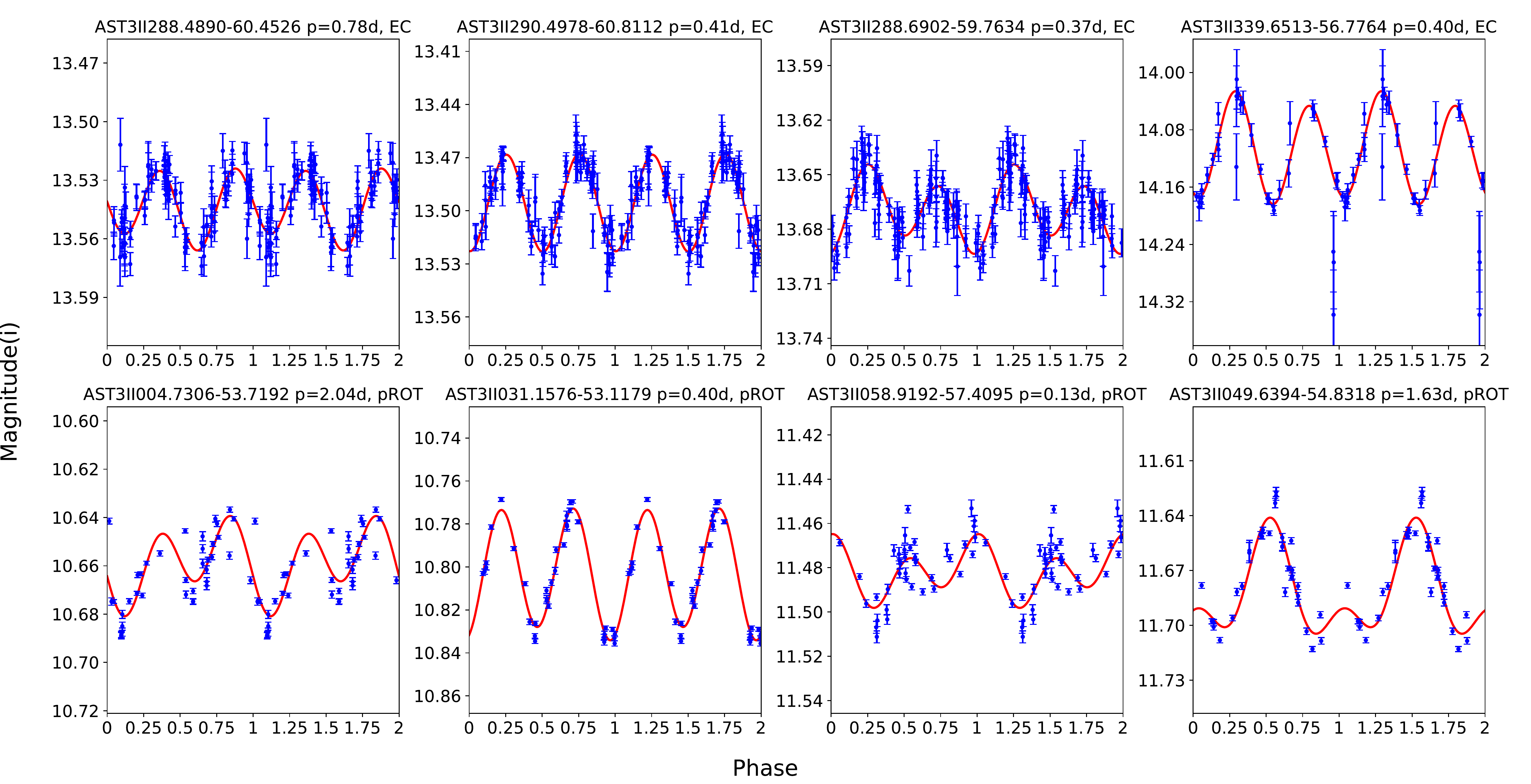}
\contcaption{Light curves of the new candidates in different classes folded in 2 phases 
except the LPVs show no apparent periods and are in the form of time series.  The 
dots with error bars are observed data and the red curve is the fitted line.  Periods 
and types are marked on the light curves.  The pulsating stars are marked as PUL and 
the possible rotational variables are marked as pROT.  
}
\label{fig:lc-new-4}
\end{figure*}

\begin{figure*}
\includegraphics[width=\textwidth]{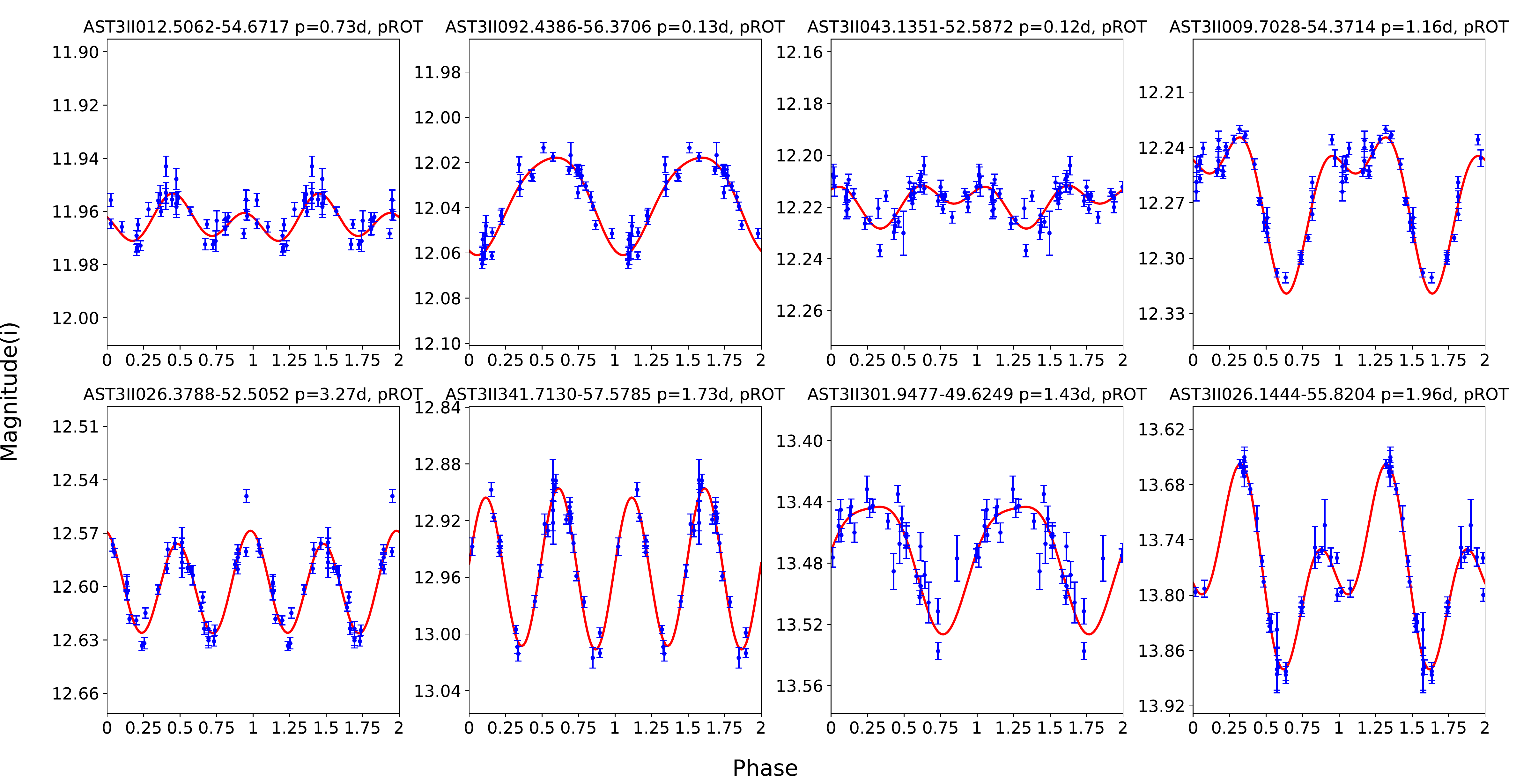}
\contcaption{Light curves of the new candidates in different classes folded in 2 phases 
except the LPVs show no apparent periods and are in the form of time series.  The 
dots with error bars are observed data and the red curve is the fitted line.  Periods 
and types are marked on the light curves.  The pulsating stars are marked as PUL and 
the possible rotational variables are marked as pROT.  
}
\label{fig:lc-new-5}
\end{figure*}

\begin{figure*}
\includegraphics[width=\textwidth]{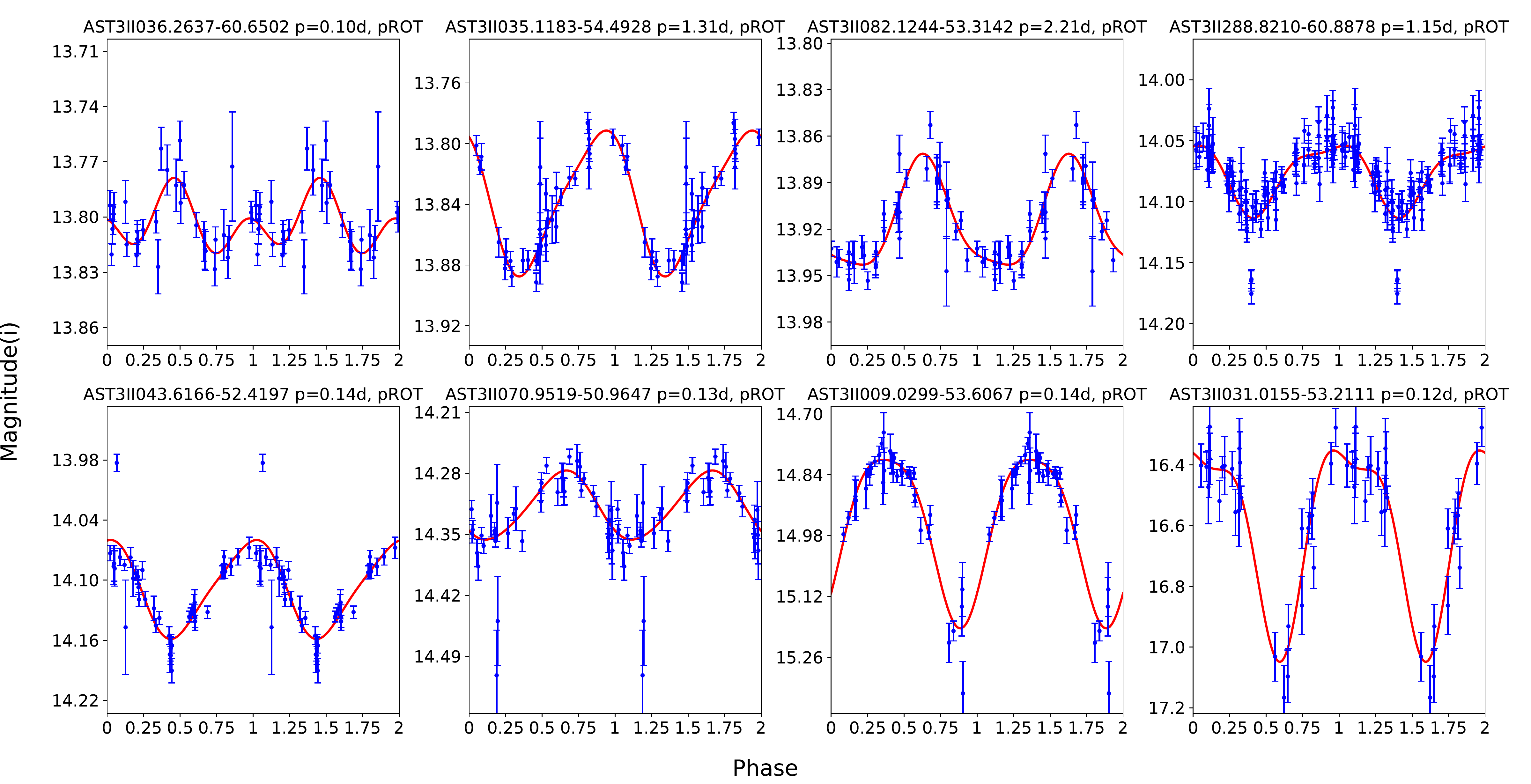}
\contcaption{Light curves of the new candidates in different classes folded in 2 phases 
except the LPVs show no apparent periods and are in the form of time series.  The 
dots with error bars are observed data and the red curve is the fitted line.  Periods 
and types are marked on the light curves.  The pulsating stars are marked as PUL and 
the possible rotational variables are marked as pROT.  
}
\label{fig:lc-new-6}
\end{figure*}

\begin{figure*}
\includegraphics[width=\textwidth]{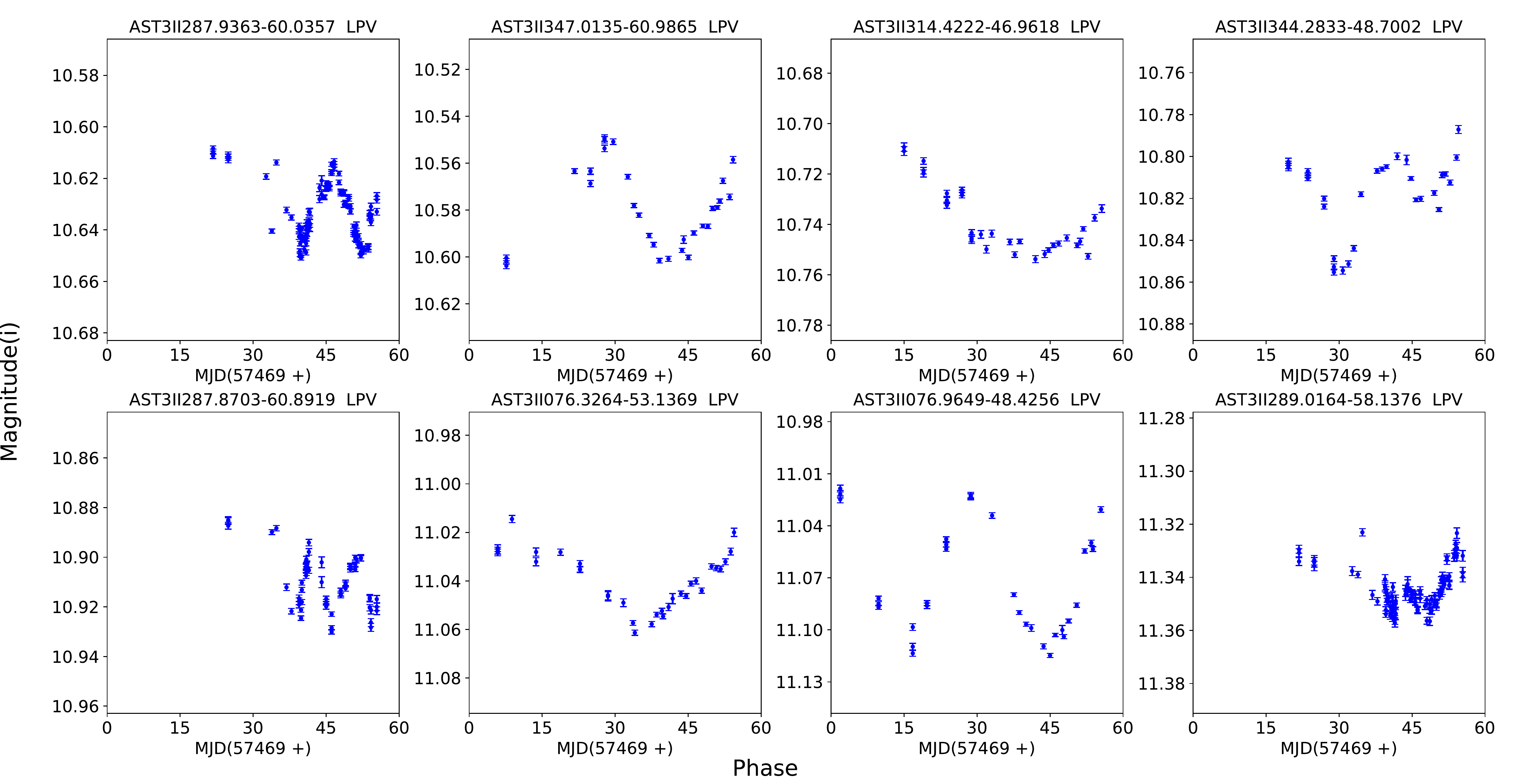}
\contcaption{Light curves of the new candidates in different classes folded in 2 phases 
except the LPVs show no apparent periods and are in the form of time series.  The 
dots with error bars are observed data and the red curve is the fitted line.  Periods 
and types are marked on the light curves.  The pulsating stars are marked as PUL and 
the possible rotational variables are marked as pROT.  
}
\label{fig:lc-new-7}
\end{figure*}

\begin{figure*}
\includegraphics[width=\textwidth]{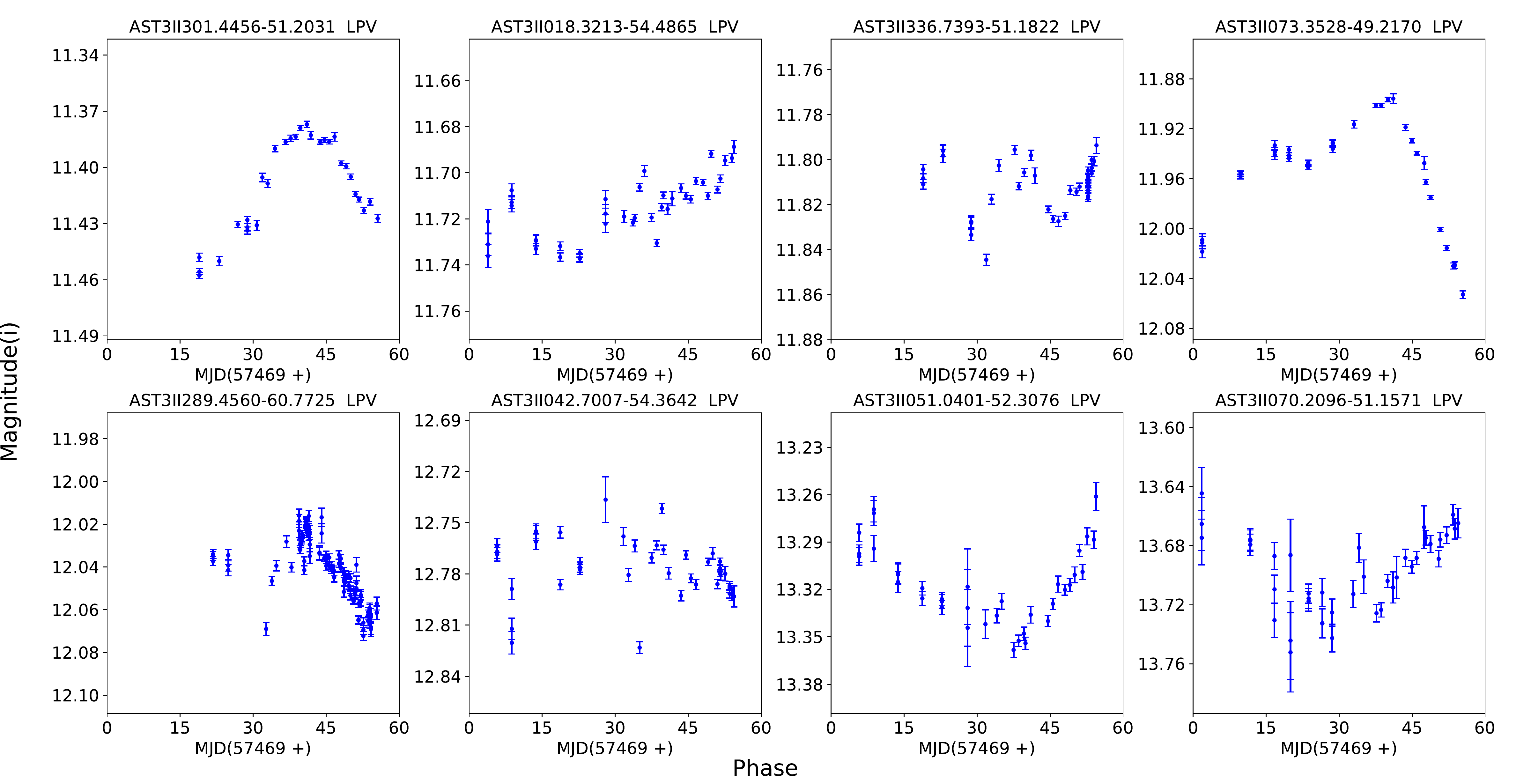}
\contcaption{Light curves of the new candidates in different classes folded in 2 phases 
except the LPVs show no apparent periods and are in the form of time series.  The 
dots with error bars are observed data and the red curve is the fitted line.  Periods 
and types are marked on the light curves.  The pulsating stars are marked as PUL and 
the possible rotational variables are marked as pROT.  
}
\label{fig:lc-new-8}
\end{figure*}

\begin{figure*}
\includegraphics[width=\textwidth]{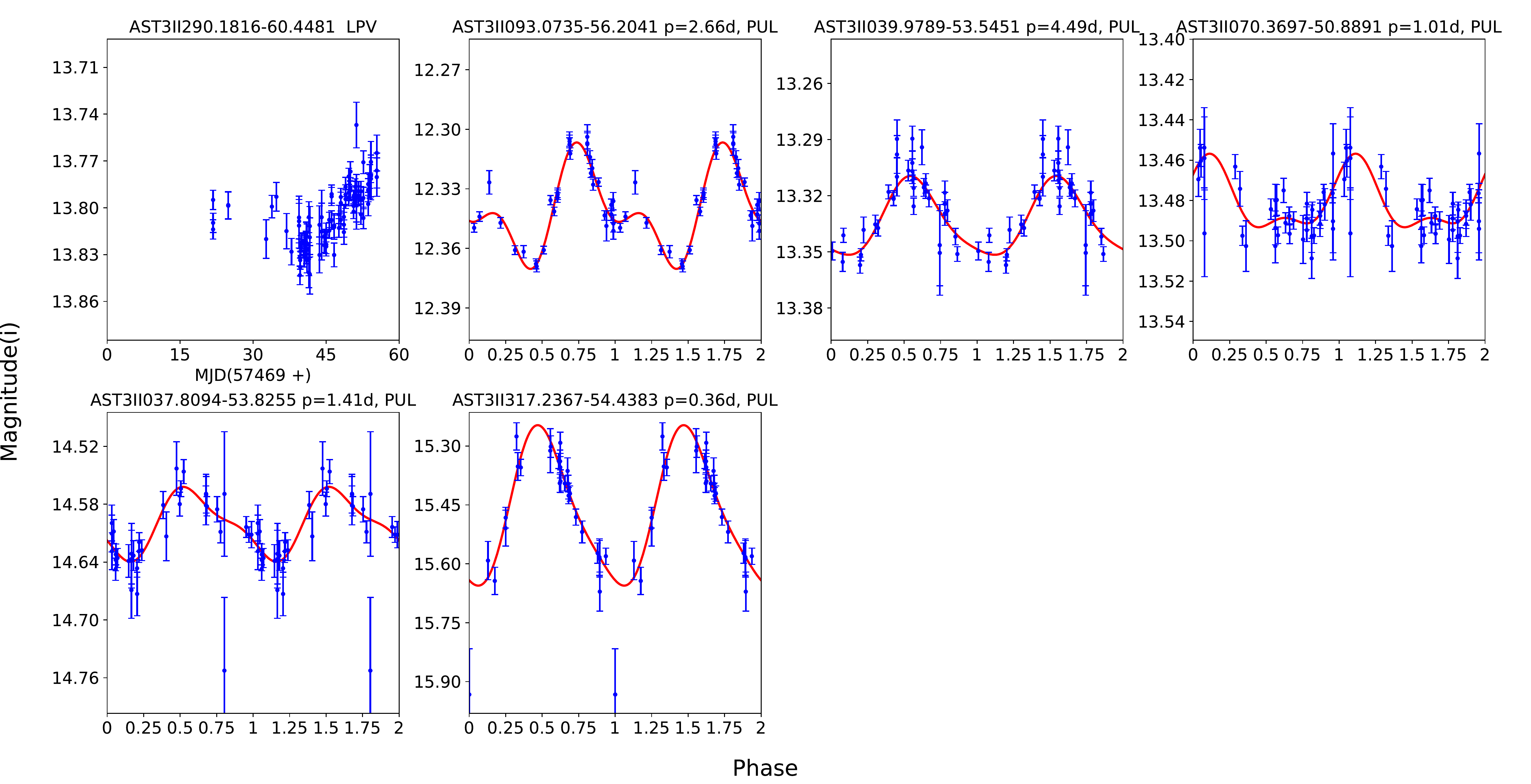}
\contcaption{Light curves of the new candidates in different classes folded in 2 phases 
except the LPVs show no apparent periods and are in the form of time series.  The 
dots with error bars are observed data and the red curve is the fitted line.  Periods 
and types are marked on the light curves.  The pulsating stars are marked as PUL and 
the possible rotational variables are marked as pROT.  
}
\label{fig:lc-new-9}
\end{figure*}

\begin{table*}
\label{tab:variablelist}
\caption{The list of new variable candidates.}
\begin{tabular}{lccccclcccc}
\hline
    Name$^{\rm a}$ & Magnitude$^{\rm b}$  & Period & Amplitude & Type & B-V & B-V ref. &
    Teff$^{\rm c}$ & logg & [Fe/H] &\cr
    & (mag) & (days) & (mag) & & (mag) & & (K) & ([cm/s$^2$]) & \\
\hline
    AST3II004.7306-53.7192 & 10.7 & 2.04 & 0.05 & ROT? & 0.22 & UCAC4$\rm ^d$ & 7714 & 3.90 & -0.38\\
    AST3II031.1576-53.1179 & 10.8 & 0.40 & 0.07 & ROT? & 0.76 & UCAC4 & 5888 & 4.09 & -0.18\\
    AST3II049.6394-54.8318 & 11.7 & 1.63 & 0.09 & ROT? & 0.52 & UCAC4 & 5489 & 3.68 & -0.81\\
    AST3II058.9192-57.4095 & 11.5 & 0.13 & 0.06 & ROT? & 0.85 & UCAC4 & 6201 & 4.49 & -0.78\\
    AST3II012.5062-54.6717 & 12 & 0.73 & 0.03 & ROT? & 1.21 & UCAC4 & 5210 & 4.48 & 0.27\\
    AST3II092.4386-56.3706 & 12 & 0.13 & 0.05 & ROT? & 0.79 & UCAC4 & 5703 & 4.33 & -0.02\\
    AST3II043.1351-52.5872 & 12.2 & 0.12 & 0.03 & ROT? & 0.53 & UCAC4 & 6257 & 4.30 & -0.19\\
    AST3II009.7028-54.3714 & 12.3 & 1.16 & 0.08 & ROT? & 0.44 & UCAC4 & 6039 & 3.94 & -0.79\\
    AST3II026.3788-52.5052 & 12.6 & 3.27 & 0.08 & ROT? & 1.55 & APASS9$\rm ^e$ & 3141 & 4.70 & 0.06\\
    AST3II341.7130-57.5785 & 12.9 & 1.73 & 0.13 & ROT? & 0.90 & UCAC4 & 5546 & 4.50 & -0.10\\
    AST3II301.9477-49.6249 & 13.5 & 1.43 & 0.11 & ROT? & 1.28 & NOMAD$\rm ^f$ & 3767 & 4.64 & 0.21\\
    AST3II026.1444-55.8204 & 13.8 & 1.96 & 0.24 & ROT? & 0.88 & UCAC4 & 5177 & 3.45 & -0.53\\
    AST3II036.2637-60.6502 & 13.8 & 0.10 & 0.07 & ROT? & 0.86 & UCAC4 & 5137 & 4.59 & -0.13\\
    AST3II035.1183-54.4928 & 13.9 & 1.31 & 0.11 & ROT? & 0.85 & UCAC4 & 5329 & 4.57 & -0.22\\
    AST3II082.1244-53.3142 & 13.9 & 2.21 & 0.10 & ROT? & 1.30 & UCAC4 & 5077 & 4.51 & 0.19\\
    AST3II043.6166-52.4197 & 14.1 & 0.14 & 0.21 & ROT? & 0.92 & UCAC4 & 5826 & 4.32 & -0.12\\
    AST3II288.8210-60.8878 & 14.1 & 1.15 & 0.15 & ROT? & 1.44 & UCAC4 & 4933 & 4.60 & 0.00\\
    AST3II070.9519-50.9647 & 14.3 & 0.13 & 0.25 & ROT? & 1.02 & UCAC4 & 5433 & 4.41 & 0.12\\
    AST3II009.0299-53.6067 & 14.8 & 0.14 & 0.60 & ROT? & 0.75 & NOMAD & 5482 & 4.40 & 0.12\\
    AST3II031.0155-53.2111 & 16.5 & 0.12 & 0.89 & ROT? & 1.71 & NOMAD & 4145 & 4.62 & 0.40\\

\hline
\end{tabular}
\flushright
\textit{Continued on next page}
\end{table*}

\begin{table*}
\label{tab:continue}
\contcaption{The list of new variable candidates.}
\begin{tabular}{lccccclcccc}
\hline
    Name$^{\rm a}$ & Mag$^{\rm b}$  & P & Amp. & Type & B-V & B-V ref. &
    Teff$^{\rm c}$ & logg & [Fe/H] &\cr
    & (mag) & (days) & (mag) & & (mag) & & (K) & ([cm/s$^2$]) & \\
\hline
    AST3II290.4978-60.8112 & 13.5 & 0.41 & 0.08 & EC & 0.60 & UCAC4 & 6397 & 3.97 & -0.29\\
    AST3II288.4890-60.4526 & 13.5 & 0.78 & 0.06 & EC & 0.43 & UCAC4 & 6349 & 2.68 & -1.38\\
    AST3II288.6902-59.7634 & 13.7 & 0.37 & 0.07 & EC & 0.85 & UCAC4 & 5801 & 4.13 & 0.00\\
    AST3II339.6513-56.7764 & 14.1 & 0.40 & 0.33 & EC & 0.28 & UCAC4 & 6630 & 4.08 & -0.49\\
    AST3II287.9363-60.0357 & 10.6 & 23.72 & 0.04 & LPV & 1.07 & UCAC4 & 4480 & 1.14 & -1.40\\
    AST3II314.4222-46.9618 & 10.7 & 200 & 0.04 & LPV & 0.30 & UCAC4 & 8526 & 4.25 & -0.15\\
    AST3II347.0135-60.9865 & 10.6 & 55.93 & 0.05 & LPV & 0.89 & UCAC4 & 5023 & 3.55 & -0.11\\
    AST3II344.2833-48.7002 & 10.8 & 34.12 & 0.07 & LPV & 1.54 & UCAC4 & 3728 & 0.79 & -0.21\\
    AST3II076.3264-53.1369 & 11 & 54.39 & 0.05 & LPV & 0.80 & UCAC4 & 5410 & 3.30 & -0.40\\
    AST3II287.8703-60.8919 & 10.9 & 16.93 & 0.05 & LPV & 1.29 & UCAC4 & 4611 & 2.41 & 0.10\\
    AST3II076.9649-48.4256 & 11.1 & 28.69 & 0.1 & LPV & 1.10 & UCAC4 & 4831 & 2.30 & -0.64\\
    AST3II289.0164-58.1376 & 11.3 & 15.52 & 0.03 & LPV & 1.53 & UCAC4 & 4007 & 0.93 & -0.58\\
    AST3II301.4456-51.2031 & 11.4 & 46.78 & 0.08 & LPV & 0.78 & NOMAD & 4115 & 1.28 & -0.39\\
    AST3II018.3213-54.4865 & 11.7 & 62.96 & 0.05 & LPV & 1.67 & APASS9 & 3018 & 4.94 & 0.24\\
    AST3II336.7393-51.1822 & 11.8 & 16.35 & 0.05 & LPV & 0.96 & UCAC4 & 5033 & 4.65 & -0.51\\
    AST3II073.3528-49.2170 & 11.9 & 65.51 & 0.16 & LPV & 0.92 & UCAC4 & 4805 & 2.58 & -0.47\\
    AST3II289.4560-60.7725 & 12 & 23.72 & 0.06 & LPV & 0.82 & UCAC4 & 5156 & 3.20 & -0.49\\
    AST3II042.7007-54.3642 & 12.8 & 98.65 & 0.09 & LPV & 0.59 & SPM4.0$\rm ^g$ & 6359 & 4.34 & -0.41\\
    AST3II051.0401-52.3076 & 13.3 & 123.73 & 0.10 & LPV & 1.53 & UCAC4 & 4372 & 0.81 & -1.55\\
    AST3II070.2096-51.1571 & 13.7 & 200 & 0.11 & LPV & 1.77 & UCAC4 & * & * & *\\
    AST3II290.1816-60.4481 & 13.8 & 43.79 & 0.10 & LPV & 0.91 & UCAC4 & 5153 & 3.45 & -0.49\\
    AST3II093.0735-56.2041 & 12.3 & 2.66 & 0.07 & PUL & 0.24 & UCAC4 & 5560 & 3.46 & -1.42\\
    AST3II039.9789-53.5451 & 13.3 & 4.49 & 0.07 & PUL & 0.77 & UCAC4 & 5880 & 4.25 & -0.14\\
    AST3II070.3697-50.8891 & 13.5 & 1.01 & 0.05 & PUL & 0.62 & UCAC4 & 10089 & 4.40 & -0.65\\
    AST3II037.8094-53.8255 & 14.6 & 1.41 & 0.21 & PUL & 1.33 & UCAC4 & 4947 & 4.65 & -0.48\\
    AST3II317.2367-54.4383 & 15.4 & 0.36 & 0.66 & PUL & 0.27 & UCAC4 & 6880 & 3.83 & -0.79\\
    AST3II041.7958-54.1112 & 11.1 & 9.19 & 0.07 & ROT & 0.50 & UCAC4 & 8240 & 4.17 & 0.05\\
    AST3II079.3367-50.0352 & 11.7 & 19.07 & 0.06 & ROT & 0.90 & UCAC4 & 5326 & 4.56 & -0.14\\
    AST3II104.4745-60.3626 & 11.8 & 8.31 & 0.07 & ROT & 1.22 & UCAC4 & 5569 & 4.56 & -0.32\\
    AST3II318.5705-51.6349 & 12.1 & 18.48 & 0.06 & ROT & 1.06 & UCAC4 & 4817 & 2.53 & -0.32\\
    AST3II024.5874-53.8901 & 12.3 & 4.7 & 0.08 & ROT & 1.55 & APASS9 & 3272 & 4.84 & 0.09\\
    AST3II084.6660-53.8947 & 12.3 & 11.72 & 0.10 & ROT & 0.88 & UCAC4 & 5543 & 3.89 & -0.37\\
    AST3II005.5981-52.7215 & 12.6 & 8.96 & 0.10 & ROT & 0.77 & UCAC4 & 5385 & 3.75 & -0.35\\
    AST3II008.2527-58.0465 & 12.7 & 15.32 & 0.06 & ROT & 0.71 & UCAC4 & 5427 & 3.53 & -0.85\\
    AST3II081.3298-46.8904 & 12.9 & 1.71 & 0.08 & ROT & 0.64 & UCAC4 & 6093 & 4.03 & -0.46\\
    AST3II025.8764-53.9192 & 13 & 3.83 & 0.04 & ROT & 0.72 & UCAC4 & 5562 & 4.41 & -0.27\\
    AST3II062.9212-52.3332 & 13.1 & 5.76 & 0.11 & ROT & 1.54 & UCAC4 & 3221 & 4.78 & 0.44\\
    AST3II084.1197-53.2223 & 13.2 & 7.42 & 0.07 & ROT & 0.76 & UCAC4 & 5156 & 3.53 & -0.36\\
    AST3II290.0774-60.2915 & 13.1 & 10.29 & 0.08 & ROT & 1.14 & UCAC4 & 6085 & 4.28 & -0.38\\
    AST3II288.7966-60.6563 & 13.2 & 12.15 & 0.07 & ROT & 1.10 & UCAC4 & 4723 & 2.53 & -0.22\\
    AST3II289.3107-60.9644 & 13.3 & 4.79 & 0.09 & ROT & 0.90 & UCAC4 & 5185 & 3.49 & -0.39\\
    AST3II073.2523-44.4335 & 13.6 & 0.13 & 0.11 & ROT & 0.58 & UCAC4 & 6146 & 4.30 & -0.31\\
    AST3II008.6057-56.4412 & 13.6 & 23.34 & 0.09 & ROT & 0.98 & UCAC4 & 4962 & 2.94 & -0.48\\
    AST3II288.1145-59.8068 & 13.5 & 3.08 & 0.07 & ROT & 0.26 & NOMAD & 4877 & 4.59 & 0.12\\
    AST3II020.2726-53.4954 & 13.8 & 11.22 & 0.10 & ROT & 0.92 & UCAC4 & 5131 & 3.70 & -0.29\\
    AST3II088.5379-52.6032 & 14.1 & 17.70 & 0.11 & ROT & 0.97 & UCAC4 & 5072 & 3.42 & -0.31\\
    AST3II036.9102-60.2014 & 14.3 & 7.84 & 0.19 & ROT & 0.72 & UCAC4 & 5488 & 3.65 & -0.90\\
    AST3II290.2670-58.6500 & 14.7 & 6.15 & 0.20 & ROT & 1.62 & UCAC4 & 4004 & 4.69 & -0.29\\
    AST3II288.3231-58.8371 & 14.9 & 12.61 & 0.26 & ROT & 0.74 & NOMAD & 4942 & 3.36 & -0.18\\
    AST3II329.6830-50.2914 & 15.1 & 3.61 & 0.33 & ROT & 0.77 & NOMAD & 5193 & 3.28 & -0.68\\
 \hline
\end{tabular}
 \flushleft
 $\rm ^a$ The IDs of AST3-2 are in the format of “AST3II+RA.+Dec.”\\
 $\rm ^b$ The mean $i$-band magnitude from AST3-2.\\
 $\rm ^c$ The Teff, B-V, and logg are from StarHorse catalogue \citep{starhorse2019}.\\
 $\rm ^d$ UCAC4 Catalogue \citep{ucac42013}\\
 $\rm ^e$ APASS9 Catalogue \citep{henden2016}\\
 $\rm ^f$ NOMAD Catalogue \citep{nomad2004}\\
 $\rm ^g$ SPM 4.0 Catalogue \citep{spm2011}\\
\end{table*}  


\bsp	
\label{lastpage}
\end{document}